\documentclass[lettersize,journal]{IEEEtran}
\IEEEoverridecommandlockouts
\usepackage{cite}
\usepackage[T1]{fontenc}
\usepackage{graphicx}
\usepackage{amssymb}
\usepackage{amsmath}
\usepackage{amsthm}
\usepackage{subcaption} 
\usepackage{microtype}
\usepackage{balance}
\usepackage{xcolor}
\usepackage{algorithm}
\usepackage{algorithmicx}
\usepackage{algpseudocode}
\usepackage[acronym]{glossaries}
\usepackage{url}
\usepackage{float}
\usepackage{booktabs}
\usepackage{hyperref}
\usepackage{graphicx}

\algrenewcommand\algorithmicindent{0.7em}%

\newacronym{NPRACH}{NPRACH}{narrowband physical random-access channel}
\newacronym{ToA}{ToA}{time of arrival}
\newacronym{CFO}{CFO}{carrier frequency offset}
\newacronym{5GNR}{5G NR}{5G New Radio}
\newacronym{3GPP}{3GPP}{3rd Generation Partnership Project}
\newacronym{UMi}{UMi}{urban microcell}
\newacronym{RMSE}{RMSE}{root-mean-square error}
\newacronym{NN}{NN}{neural network}
\newacronym{BS}{BS}{base station}
\newacronym{UE}{UE}{user equipment}
\newacronym{CP}{CP}{cyclic prefix}
\newacronym{OFDM}{OFDM}{orthogonal frequency division multiplexing}
\newacronym{FFT}{FFT}{fast Fourier transform}
\newacronym{AWGN}{AWGN}{additive white Gaussian noise}
\newacronym{DFT}{DFT}{discrete Fourier transform}
\newacronym{WSSUS}{WSSUS}{wide-sense stationary uncorrelated scattering}
\newacronym{PDP}{PDP}{power delay profile}
\newacronym{RG}{RG}{resource grid}
\newacronym{RE}{RE}{resource element}
\newacronym{SNR}{SNR}{signal-to-noise ratio}
\newacronym{SINR}{SINR}{signal-to-interference-plus-noise ratio}
\newacronym{MLP}{MLP}{multilayer perceptron}
\newacronym{BCE}{BCE}{binary cross-entropy}
\newacronym{CCE}{CCE}{categorical cross-entropy}
\newacronym{ERM}{ERM}{empirical risk minimization}
\newacronym{MC}{MC}{Monte Carlo}
\newacronym{KL}{KL}{Kullback–Leibler}
\newacronym{SGD}{SGD}{stochastic gradient descent}
\newacronym{ICI}{ICI}{inter-carrier interference}
\newacronym{GNN}{GNN}{graph neural network}
\newacronym{BP}{BP}{belief propagation}
\newacronym{MP}{MP}{message passing}
\newacronym{FEC}{FEC}{forward error correction}
\newacronym{LDPC}{LDPC}{low-density parity-check}
\newacronym{IQ}{I/Q}{in-phase/quadrature}
\newacronym{HDPC}{HDPC}{high-density parity-check}
\newacronym{SCL}{SCL}{successive cancellation list}
\newacronym{SC}{SC}{successive cancellation}
\newacronym{URLLC}{URLLC}{ultra-reliable low-latency communications}
\newacronym{APP}{APP}{a posterior probability}
\newacronym{MIMO}{MIMO}{multiple-input multiple-output}
\newacronym{CNN}{CNN}{convolutional neural network}
\newacronym{ResNet}{ResNet}{residual network}
\newacronym{BER}{BER}{bit error rate}
\newacronym{BPSK}{BPSK}{binary phase shift keying}
\newacronym{LLR}{LLR}{log-likelihood ratio}
\newacronym{VN}{VN}{variable node}
\newacronym{CN}{CN}{check node}
\newacronym{MPNN}{MPNN}{message passing neural network}

\newacronym{AI}{AI}{artificial intelligence}
\newacronym{ML}{ML}{machine learning}
\newacronym{GMM}{GMM}{Gaussian mixture model}
\newacronym{LLM}{LLM}{large language model}
\newacronym{SISO}{SISO}{single-input single-output; soft-input soft-output}
\newacronym{MISO}{MISO}{multiple-input single-output}
\newacronym{PRB}{PRB}{physical resource block}
\newacronym{PUSCH}{PUSCH}{physical uplink shared channel}
\newacronym{PDSCH}{PDSCH}{physical downlink shared channel}
\newacronym{PUCCH}{PUCCH}{physical uplink control channel}
\newacronym{PDCCH}{PDCCH}{physical downlink control channel}
\newacronym{PRACH}{PRACH}{physical random-access channel}
\newacronym{SRS}{SRS}{sounding reference signal}
\newacronym{PBCH}{PBCH}{physical broadcast channel}
\newacronym{SSB}{SSB}{synchronization signal block}
\newacronym{AMF}{AMF}{access and mobility management function}
\newacronym{SMF}{SMF}{session management function}
\newacronym{PDU}{PDU}{protocol data unit}
\newacronym{NGAP}{NGAP}{next generation application protocol}
\newacronym{IMS}{IMS}{IP multimedia subsystem}
\newacronym{IP}{IP}{Internet Protocol}
\newacronym{SIP}{SIP}{session initiation protocol}
\newacronym{DN}{DN}{data network}
\newacronym{UPF}{UPF}{user plane function}
\newacronym{AUSF}{AUSF}{authentication server function}
\newacronym{UDM}{UDM}{unified data management}
\newacronym{UDR}{UDR}{unified data repository}
\newacronym{NRF}{NRF}{network repository function}
\newacronym{HARQ}{HARQ}{hybrid automatic repeat request}
\newacronym{LA}{LA}{link adaptation}
\newacronym{SE}{SE}{spectral efficiency}
\newacronym{TCP}{TCP}{transmission control protocol}
\newacronym{API}{API}{application programming interface}
\newacronym{SQL}{SQL}{structured query language}
\newacronym{ARF}{ARF}{auto rate fallback}
\newacronym{ACK}{ACK}{acknowledgment}
\newacronym{NACK}{NACK}{negative acknowledgment}
\newacronym{OLLA}{OLLA}{outer-loop link adaptation}
\newacronym{CQI}{CQI}{channel quality indicator}
\newacronym{RL}{RL}{reinforcement learning}
\newacronym{MAB}{MAB}{multi-armed bandit}
\newacronym{SALAD}{SALAD}{self-adaptive link adaptation}
\newacronym{ILLA}{ILLA}{inner-loop link adaptation}
\newacronym{OGD}{OGD}{online gradient descent}

\newacronym{MUMIMO}{MU-MIMO}{multi-user multiple-input multiple-output}
\newacronym{ULL}{ULL}{uplink layer}
\newacronym{BICM}{BICM}{bit-interleaved coded modulation}
\newacronym{QAM}{QAM}{quadrature amplitude modulation}
\newacronym{LMMSE}{LMMSE}{linear minimum mean square error}
\newacronym{MMSE}{MMSE}{minimum mean square error}
\newacronym{CSI}{CSI}{channel-state information}
\newacronym{SIMO}{SIMO}{single-input multiple-output}
\newacronym{SLAM}{SLAM}{simultaneous localization and mapping}
\newacronym{LIDAR}{LIDAR}{light detection and ranging}
\newacronym{CGNN}{CGNN}{convolutional and graph neural network}
\newacronym{BLER}{BLER}{block error rate}
\newacronym{LS}{LS}{least squares}
\newacronym{PE}{PE}{positional encoding}
\newacronym{relu}{ReLU}{rectified linear unit}
\newacronym{RB}{RB}{resource block}
\newacronym{DMRS}{DMRS}{demodulation reference signal}
\newacronym{CSI-RS}{CSI-RS}{channel state information reference signal}
\newacronym{IoT}{IoT}{internet of things}
\newacronym{ADAM}{ADAM}{adaptive momentum}
\newacronym{TBLER}{TBLER}{transport block error rate}
\newacronym{TBS}{TBS}{transport block size}
\newacronym{MCS}{MCS}{modulation and coding scheme}
\newacronym{TDL}{TDL}{tapped delay line}
\newacronym{CDL}{CDL}{clustered delay line}
\newacronym{GSCM}{GSCM}{geometry-based stochastic channel model}
\newacronym{CDM}{CDM}{code division multiplexing}
\newacronym{FLOP}{FLOP}{floating point operation}
\newacronym{PHY}{PHY}{physical layer}
\newacronym{MAC}{MAC}{media access control}
\newacronym{gNB}{gNB}{next-generation Node B}
\newacronym{RLC}{RLC}{radio link control}
\newacronym{PDCP}{PDCP}{packet data convergence protocol}
\newacronym{SDAP}{SDAP}{service data adaptation protocol}
\newacronym{RRC}{RRC}{radio resource control}

\newacronym{ULA}{ULA}{uniform linear array}
\newacronym{NRX}{NRX}{neural receiver}
\newacronym{MDX}{MDX}{model-driven neural receiver}
\newacronym{GPU}{GPU}{graphics processing unit}
\newacronym{CPU}{CPU}{central processing unit}
\newacronym{NIC}{NIC}{network interface card}
\newacronym{Var-MCS-NRX}{Var-MCS NRX}{variable-MCS NRX}
\newacronym{UL}{UL}{uplink}
\newacronym{DL}{DL}{downlink}
\newacronym{MSE}{MSE}{mean squared error}
\newacronym{CIR}{CIR}{channel impulse response}
\newacronym{iid}{iid}{independent and identically distributed}

\newacronym{RAN}{RAN}{radio access network}
\newacronym{ORU}{O-RU}{open RAN radio unit}
\newacronym{ORAN}{O-RAN}{open radio access network}
\newacronym{COTS}{COTS}{commercial-off-the-shelf}
\newacronym{RF}{RF}{radio frequency}
\newacronym{LOS}{LOS}{line-of-sight}
\newacronym{NLOS}{NLOS}{non-line-of-sight}
\newacronym{OTA}{OTA}{over-the-air}
\newacronym{TDD}{TDD}{time-division duplexing}
\newacronym{CAEZ}{CAEZ}{CSI acquisition at ETH Zurich}
\newacronym{ATB}{ATB}{NVIDIA Aerial Testbed}
\newacronym{ARK}{ARK}{NVIDIA Aerial Research Kit}
\newacronym{OAI}{OAI}{OpenAirInterface}
\newacronym{RFFI}{RFFI}{radio frequency fingerprint identification}
\newacronym{FH}{FH}{front haul}
\newacronym{GNSS}{GNSS}{global navigation satellite system}
\newacronym{RTK}{RTK}{real-time kinematic}
\newacronym{PTP}{PTP}{precision time protocol}
\newacronym{RNTI}{RNTI}{radio network temporary identifier}
\newacronym{ISM}{ISM}{Industrial, Scientific, and Medical}
\newacronym{UAV}{UAV}{unmanned aerial vehicle}
\newacronym{CDF}{CDF}{cumulative distribution function}
\newacronym{CRC}{CRC}{cyclic redundancy check}
\newacronym{TP}{TP}{throughput}

\newacronym{IDD}{IDD}{iterative detection and decoding}
\newacronym{DUIDD}{DUIDD}{deep-unfolded interleaved detection and decoding}
\newacronym{SPAT}{SPAT}{swapping of punctured and transmitted blocks}
\newacronym{SDR}{SDR}{software-defined radio}
\usepackage{amssymb}
\usepackage{amsfonts}
\usepackage{mathrsfs}
\usepackage{xspace}
\usepackage{bm}
\usepackage{upgreek}

\newcommand{\safemath}[2]{\newcommand{#1}{\ensuremath{#2}\xspace}}

\safemath{\bma}{\mathbf{a}}
\safemath{\bmb}{\mathbf{b}}
\safemath{\bmc}{\mathbf{c}}
\safemath{\bmd}{\mathbf{d}}
\safemath{\bme}{\mathbf{e}}
\safemath{\bmf}{\mathbf{f}}
\safemath{\bmg}{\mathbf{g}}
\safemath{\bmh}{\mathbf{h}}
\safemath{\bmi}{\mathbf{i}}
\safemath{\bmj}{\mathbf{j}}
\safemath{\bmk}{\mathbf{k}}
\safemath{\bml}{\mathbf{l}}
\safemath{\bmm}{\mathbf{m}}
\safemath{\bmn}{\mathbf{n}}
\safemath{\bmo}{\mathbf{o}}
\safemath{\bmp}{\mathbf{p}}
\safemath{\bmq}{\mathbf{q}}
\safemath{\bmr}{\mathbf{r}}
\safemath{\bms}{\mathbf{s}}
\safemath{\bmt}{\mathbf{t}}
\safemath{\bmu}{\mathbf{u}}
\safemath{\bmv}{\mathbf{v}}
\safemath{\bmw}{\mathbf{w}}
\safemath{\bmx}{\mathbf{x}}
\safemath{\bmy}{\mathbf{y}}
\safemath{\bmz}{\mathbf{z}}
\safemath{\bmzero}{\mathbf{0}}
\safemath{\bmone}{\mathbf{1}}
\safemath{\Bell}{\ensuremath{\boldsymbol\ell}}

\bmdefine{\biad}{a}
\bmdefine{\bibd}{b}
\bmdefine{\bicd}{c}
\bmdefine{\bidd}{d}
\bmdefine{\bied}{e}
\bmdefine{\bifd}{f}
\bmdefine{\bigd}{g}
\bmdefine{\bihd}{h}
\bmdefine{\biid}{i}
\bmdefine{\bijd}{j}
\bmdefine{\bikd}{k}
\bmdefine{\bild}{l}
\bmdefine{\bimd}{m}
\bmdefine{\bind}{n}
\bmdefine{\biod}{o}
\bmdefine{\bipd}{p}
\bmdefine{\biqd}{q}
\bmdefine{\bird}{r}
\bmdefine{\bisd}{s}
\bmdefine{\bitd}{t}
\bmdefine{\biud}{u}
\bmdefine{\bivd}{v}
\bmdefine{\biwd}{w}
\bmdefine{\bixd}{x}
\bmdefine{\biyd}{y}
\bmdefine{\bizd}{z}

\bmdefine{\bixid}{\xi}
\bmdefine{\bilambdad}{\lambda}
\bmdefine{\bimud}{\mu}
\bmdefine{\bithetad}{\theta}
\bmdefine{\biphid}{\phi}
\bmdefine{\bideltad}{\delta}

\safemath{\bmia}{\biad}
\safemath{\bmib}{\bibd}
\safemath{\bmic}{\bicd}
\safemath{\bmid}{\bidd}
\safemath{\bmie}{\bied}
\safemath{\bmif}{\bifd}
\safemath{\bmig}{\bigd}
\safemath{\bmih}{\bihd}
\safemath{\bmii}{\biid}
\safemath{\bmij}{\bijd}
\safemath{\bmik}{\bikd}
\safemath{\bmil}{\bild}
\safemath{\bmim}{\bimd}
\safemath{\bmin}{\bind}
\safemath{\bmio}{\biod}
\safemath{\bmip}{\bipd}
\safemath{\bmiq}{\biqd}
\safemath{\bmir}{\bird}
\safemath{\bmis}{\bisd}
\safemath{\bmit}{\bitd}
\safemath{\bmiu}{\biud}
\safemath{\bmiv}{\bivd}
\safemath{\bmiw}{\biwd}
\safemath{\bmix}{\bixd}
\safemath{\bmiy}{\biyd}
\safemath{\bmiz}{\bizd}

\safemath{\bmxi}{\bixid}
\safemath{\bmlambda}{\bilambdad}
\safemath{\bmmu}{\bimud}
\safemath{\bmtheta}{\bithetad}
\safemath{\bmphi}{\biphid}
\safemath{\bmdelta}{\bideltad}

\safemath{\bA}{\mathbf{A}}
\safemath{\bB}{\mathbf{B}}
\safemath{\bC}{\mathbf{C}}
\safemath{\bD}{\mathbf{D}}
\safemath{\bE}{\mathbf{E}}
\safemath{\bF}{\mathbf{F}}
\safemath{\bG}{\mathbf{G}}
\safemath{\bH}{\mathbf{H}}
\safemath{\bI}{\mathbf{I}}
\safemath{\bJ}{\mathbf{J}}
\safemath{\bK}{\mathbf{K}}
\safemath{\bL}{\mathbf{L}}
\safemath{\bM}{\mathbf{M}}
\safemath{\bN}{\mathbf{N}}
\safemath{\bO}{\mathbf{O}}
\safemath{\bP}{\mathbf{P}}
\safemath{\bQ}{\mathbf{Q}}
\safemath{\bR}{\mathbf{R}}
\safemath{\bS}{\mathbf{S}}
\safemath{\bT}{\mathbf{T}}
\safemath{\bU}{\mathbf{U}}
\safemath{\bV}{\mathbf{V}}
\safemath{\bW}{\mathbf{W}}
\safemath{\bX}{\mathbf{X}}
\safemath{\bY}{\mathbf{Y}}
\safemath{\bZ}{\mathbf{Z}}

\safemath{\bZero}{\mathbf{0}}
\safemath{\bOne}{\mathbf{1}}
\safemath{\bDelta}{\mathbf{\Delta}}
\safemath{\bLambda}{\mathbf{\UpLambda}}
\safemath{\bPhi}{\mathbf{\Upphi}}
\safemath{\bSigma}{\mathbf{\Upsigma}}
\safemath{\bOmega}{\mathbf{\Upomega}}
\safemath{\bTheta}{\mathbf{\Uptheta}}

\bmdefine{\biAd}{A}
\bmdefine{\biBd}{B}
\bmdefine{\biCd}{C}
\bmdefine{\biDd}{D}
\bmdefine{\biEd}{E}
\bmdefine{\biFd}{F}
\bmdefine{\biGd}{G}
\bmdefine{\biHd}{H}
\bmdefine{\biId}{I}
\bmdefine{\biJd}{J}
\bmdefine{\biKd}{K}
\bmdefine{\biLd}{L}
\bmdefine{\biMd}{M}
\bmdefine{\biOd}{N}
\bmdefine{\biPd}{O}
\bmdefine{\biQd}{P}
\bmdefine{\biRd}{R}
\bmdefine{\biSd}{S}
\bmdefine{\biTd}{T}
\bmdefine{\biUd}{U}
\bmdefine{\biVd}{V}
\bmdefine{\biWd}{W}
\bmdefine{\biXd}{X}
\bmdefine{\biYd}{Y}
\bmdefine{\biZd}{Z}

\bmdefine{\biDelta}{\Delta}
\bmdefine{\biLambda}{\Lambda}
\bmdefine{\biPhi}{\Phi}
\bmdefine{\biSigma}{\Sigma}
\bmdefine{\biOmega}{\Omega}
\bmdefine{\biTheta}{\Theta}

\safemath{\bimA}{\biAd}
\safemath{\bimB}{\biBd}
\safemath{\bimC}{\biCd}
\safemath{\bimD}{\biDd}
\safemath{\bimE}{\biEd}
\safemath{\bimF}{\biFd}
\safemath{\bimG}{\biGd}
\safemath{\bimH}{\biHd}
\safemath{\bimI}{\biId}
\safemath{\bimJ}{\biJd}
\safemath{\bimK}{\biKd}
\safemath{\bimL}{\biLd}
\safemath{\bimM}{\biMd}
\safemath{\bimN}{\biNd}
\safemath{\bimO}{\biOd}
\safemath{\bimP}{\biPd}
\safemath{\bimQ}{\biQd}
\safemath{\bimR}{\biRd}
\safemath{\bimS}{\biSd}
\safemath{\bimT}{\biTd}
\safemath{\bimU}{\biUd}
\safemath{\bimV}{\biVd}
\safemath{\bimW}{\biWd}
\safemath{\bimX}{\biXd}
\safemath{\bimY}{\biYd}
\safemath{\bimZ}{\biZd}

\safemath{\bimDelta}{\biDelta}
\safemath{\bimLambda}{\biLambda}
\safemath{\bimPhi}{\biPhi}
\safemath{\bimSigma}{\biSigma}
\safemath{\bimOmega}{\biOmega}
\safemath{\bimTheta}{\biTheta}

\safemath{\setA}{\mathcal{A}}
\safemath{\setB}{\mathcal{B}}
\safemath{\setC}{\mathcal{C}}
\safemath{\setD}{\mathcal{D}}
\safemath{\setE}{\mathcal{E}}
\safemath{\setF}{\mathcal{F}}
\safemath{\setG}{\mathcal{G}}
\safemath{\setH}{\mathcal{H}}
\safemath{\setI}{\mathcal{I}}
\safemath{\setJ}{\mathcal{J}}
\safemath{\setK}{\mathcal{K}}
\safemath{\setL}{\mathcal{L}}
\safemath{\setM}{\mathcal{M}}
\safemath{\setN}{\mathcal{N}}
\safemath{\setO}{\mathcal{O}}
\safemath{\setP}{\mathcal{P}}
\safemath{\setQ}{\mathcal{Q}}
\safemath{\setR}{\mathcal{R}}
\safemath{\setS}{\mathcal{S}}
\safemath{\setT}{\mathcal{T}}
\safemath{\setU}{\mathcal{U}}
\safemath{\setV}{\mathcal{V}}
\safemath{\setW}{\mathcal{W}}
\safemath{\setX}{\mathcal{X}}
\safemath{\setY}{\mathcal{Y}}
\safemath{\setZ}{\mathcal{Z}}
\safemath{\emptySet}{\varnothing}

\safemath{\colA}{\mathscr{A}}
\safemath{\colB}{\mathscr{B}}
\safemath{\colC}{\mathscr{C}}
\safemath{\colD}{\mathscr{D}}
\safemath{\colE}{\mathscr{E}}
\safemath{\colF}{\mathscr{F}}
\safemath{\colG}{\mathscr{G}}
\safemath{\colH}{\mathscr{H}}
\safemath{\colI}{\mathscr{I}}
\safemath{\colJ}{\mathscr{J}}
\safemath{\colK}{\mathscr{K}}
\safemath{\colL}{\mathscr{L}}
\safemath{\colM}{\mathscr{M}}
\safemath{\colN}{\mathscr{N}}
\safemath{\colO}{\mathscr{O}}
\safemath{\colP}{\mathscr{P}}
\safemath{\colQ}{\mathscr{Q}}
\safemath{\colR}{\mathscr{R}}
\safemath{\colS}{\mathscr{S}}
\safemath{\colT}{\mathscr{T}}
\safemath{\colU}{\mathscr{U}}
\safemath{\colV}{\mathscr{V}}
\safemath{\colW}{\mathscr{W}}
\safemath{\colX}{\mathscr{X}}
\safemath{\colY}{\mathscr{Y}}
\safemath{\colZ}{\mathscr{Z}}

\safemath{\opA}{\mathbb{A}}
\safemath{\opB}{\mathbb{B}}
\safemath{\opC}{\mathbb{C}}
\safemath{\opD}{\mathbb{D}}
\safemath{\opE}{\mathbb{E}}
\safemath{\opF}{\mathbb{F}}
\safemath{\opG}{\mathbb{G}}
\safemath{\opH}{\mathbb{H}}
\safemath{\opI}{\mathbb{I}}
\safemath{\opJ}{\mathbb{J}}
\safemath{\opK}{\mathbb{K}}
\safemath{\opL}{\mathbb{L}}
\safemath{\opM}{\mathbb{M}}
\safemath{\opN}{\mathbb{N}}
\safemath{\opO}{\mathbb{O}}
\safemath{\opP}{\mathbb{P}}
\safemath{\opQ}{\mathbb{Q}}
\safemath{\opR}{\mathbb{R}}
\safemath{\opS}{\mathbb{S}}
\safemath{\opT}{\mathbb{T}}
\safemath{\opU}{\mathbb{U}}
\safemath{\opV}{\mathbb{V}}
\safemath{\opW}{\mathbb{W}}
\safemath{\opX}{\mathbb{X}}
\safemath{\opY}{\mathbb{Y}}
\safemath{\opZ}{\mathbb{Z}}
\safemath{\opZero}{\mathbb{O}}
\safemath{\identityop}{\opI}

\safemath{\veca}{\bma}
\safemath{\vecb}{\bmb}
\safemath{\vecc}{\bmc}
\safemath{\vecd}{\bmd}
\safemath{\vece}{\bme}
\safemath{\vecf}{\bmf}
\safemath{\vecg}{\bmg}
\safemath{\vech}{\bmh}
\safemath{\veci}{\bmi}
\safemath{\vecj}{\bmj}
\safemath{\veck}{\bmk}
\safemath{\vecl}{\bml}
\safemath{\vecm}{\bmm}
\safemath{\vecn}{\bmn}
\safemath{\veco}{\bmo}
\safemath{\vecp}{\bmp}
\safemath{\vecq}{\bmq}
\safemath{\vecr}{\bmr}
\safemath{\vecs}{\bms}
\safemath{\vect}{\bmt}
\safemath{\vecu}{\bmu}
\safemath{\vecv}{\bmv}
\safemath{\vecw}{\bmw}
\safemath{\vecx}{\bmx}
\safemath{\vecy}{\bmy}
\safemath{\vecz}{\bmz}

\safemath{\veczero}{\bmzero}
\safemath{\vecone}{\bmone}
\safemath{\vecxi}{\bmxi}
\safemath{\veclambda}{\bmlambda}
\safemath{\vecmu}{\bmmu}
\safemath{\vectheta}{\bmtheta}
\safemath{\vecphi}{\bmphi}
\safemath{\vecdelta}{\bmdelta}

\safemath{\matA}{\bA}
\safemath{\matB}{\bB}
\safemath{\matC}{\bC}
\safemath{\matD}{\bD}
\safemath{\matE}{\bE}
\safemath{\matF}{\bF}
\safemath{\matG}{\bG}
\safemath{\matH}{\bH}
\safemath{\matI}{\bI}
\safemath{\matJ}{\bJ}
\safemath{\matK}{\bK}
\safemath{\matL}{\bL}
\safemath{\matM}{\bM}
\safemath{\matN}{\bN}
\safemath{\matO}{\bO}
\safemath{\matP}{\bP}
\safemath{\matQ}{\bQ}
\safemath{\matR}{\bR}
\safemath{\matS}{\bS}
\safemath{\matT}{\bT}
\safemath{\matU}{\bU}
\safemath{\matV}{\bV}
\safemath{\matW}{\bW}
\safemath{\matX}{\bX}
\safemath{\matY}{\bY}
\safemath{\matZ}{\bZ}
\safemath{\matzero}{\bmzero}

\safemath{\matDelta}{\bDelta}
\safemath{\matLambda}{\bLambda}
\safemath{\matPhi}{\bPhi}
\safemath{\matSigma}{\bSigma}
\safemath{\matOmega}{\bOmega}
\safemath{\matTheta}{\bTheta}

\safemath{\matidentity}{\matI}
\safemath{\matone}{\matO}

\safemath{\rnda}{A}
\safemath{\rndb}{B}
\safemath{\rndc}{C}
\safemath{\rndd}{D}
\safemath{\rnde}{E}
\safemath{\rndf}{F}
\safemath{\rndg}{G}
\safemath{\rndh}{H}
\safemath{\rndi}{I}
\safemath{\rndj}{J}
\safemath{\rndk}{K}
\safemath{\rndl}{L}
\safemath{\rndm}{M}
\safemath{\rndn}{N}
\safemath{\rndo}{O}
\safemath{\rndp}{P}
\safemath{\rndq}{Q}
\safemath{\rndr}{R}
\safemath{\rnds}{S}
\safemath{\rndt}{T}
\safemath{\rndu}{U}
\safemath{\rndv}{V}
\safemath{\rndw}{W}
\safemath{\rndx}{X}
\safemath{\rndy}{Y}
\safemath{\rndz}{Z}

\safemath{\rveca}{\bimA}
\safemath{\rvecb}{\bimB}
\safemath{\rvecc}{\bimC}
\safemath{\rvecd}{\bimD}
\safemath{\rvece}{\bimE}
\safemath{\rvecf}{\bimF}
\safemath{\rvecg}{\bimG}
\safemath{\rvech}{\bimH}
\safemath{\rveci}{\bimI}
\safemath{\rvecj}{\bimJ}
\safemath{\rveck}{\bimK}
\safemath{\rvecl}{\bimL}
\safemath{\rvecm}{\bimM}
\safemath{\rvecn}{\bimN}
\safemath{\rveco}{\bomO}
\safemath{\rvecp}{\bimP}
\safemath{\rvecq}{\bimQ}
\safemath{\rvecr}{\bimR}
\safemath{\rvecs}{\bimS}
\safemath{\rvect}{\bimT}
\safemath{\rvecu}{\bimU}
\safemath{\rvecv}{\bimV}
\safemath{\rvecw}{\bimW}
\safemath{\rvecx}{\bimX}
\safemath{\rvecy}{\bimY}
\safemath{\rvecz}{\bimZ}

\safemath{\rvecxi}{\bmxi}
\safemath{\rveclambda}{\bmlambda}
\safemath{\rvecmu}{\bmmu}
\safemath{\rvectheta}{\bmtheta}
\safemath{\rvecphi}{\bmphi}

\safemath{\rmatA}{\bimA}
\safemath{\rmatB}{\bimB}
\safemath{\rmatC}{\bimC}
\safemath{\rmatD}{\bimD}
\safemath{\rmatE}{\bimE}
\safemath{\rmatF}{\bimF}
\safemath{\rmatG}{\bimG}
\safemath{\rmatH}{\bimH}
\safemath{\rmatI}{\bimI}
\safemath{\rmatJ}{\bimJ}
\safemath{\rmatK}{\bimK}
\safemath{\rmatL}{\bimL}
\safemath{\rmatM}{\bimM}
\safemath{\rmatN}{\bimN}
\safemath{\rmatO}{\bimO}
\safemath{\rmatP}{\bimP}
\safemath{\rmatQ}{\bimQ}
\safemath{\rmatR}{\bimR}
\safemath{\rmatS}{\bimS}
\safemath{\rmatT}{\bimT}
\safemath{\rmatU}{\bimU}
\safemath{\rmatV}{\bimV}
\safemath{\rmatW}{\bimW}
\safemath{\rmatX}{\bimX}
\safemath{\rmatY}{\bimY}
\safemath{\rmatZ}{\bimZ}

\safemath{\rmatDelta}{\bimDelta}
\safemath{\rmatLambda}{\bimLambda}
\safemath{\rmatPhi}{\bimPhi}
\safemath{\rmatSigma}{\bimSigma}
\safemath{\rmatOmega}{\bimOmega}
\safemath{\rmatTheta}{\bimTheta}

\usepackage{amssymb}
\usepackage{amsfonts}
\usepackage{mathrsfs}
\usepackage{xspace}
\usepackage{bm}
\usepackage{fancyref}
\usepackage{textcomp}

\usepackage{multirow}
\usepackage{stmaryrd}

\newenvironment{textbmatrix}{	\setlength{\arraycolsep}{2.5pt}%
								\left[\begin{matrix}}{\end{matrix}\right]%
								\raisebox{0.08ex}{\vphantom{M}}}

\def\be{\begin{equation}}
\def\ee{\end{equation}}
\def\een{\nonumber \end{equation}}
\def\mat{\begin{bmatrix}}
\def\emat{\end{bmatrix}}
\def\btm{\begin{textbmatrix}}
\def\etm{\end{textbmatrix}}

\def\ba#1\ea{\begin{align}#1\end{align}}
\def\bas#1\eas{\begin{align*}#1\end{align*}}
\def\bs#1\es{\begin{split}#1\end{split}}
\def\bg#1\eg{\begin{gather}#1\end{gather}}
\def\bml#1\eml{\begin{multline}#1\end{multline}}
\def\bi#1\ei{\begin{itemize}#1\end{itemize}}

\newcommand{\lefto}{\mathopen{}\left}

\DeclareMathOperator{\Exop}{\opE}			

\newcommand{\Ex}[2]{\ensuremath{\Exop_{#1}\lefto[#2\right]}} 	

\newcommand{\conj}[1]{\ensuremath{#1^{*}}} 	
\newcommand{\tp}[1]{\ensuremath{#1^{T}}} 		
\newcommand{\herm}[1]{\ensuremath{#1^{H}}} 	

\safemath{\dirac}{\delta}					
\safemath{\krond}{\dirac}					

\safemath{\upto}{\uparrow}
\safemath{\downto}{\downarrow}
\safemath{\iu}{j}							
\safemath{\ev}{\lambda}						
\safemath{\hilseqspace}{l^{2}}				
\newcommand{\banachfunspace}[1]{\setL^{#1}}	
\safemath{\hilfunspace}{\banachfunspace{2}}	

\safemath{\SNR}{\textit{SNR}} 				
\safemath{\PAR}{\textit{PAR}} 				
\safemath{\No}{N_0}							
\safemath{\Es}{E_s}							
\safemath{\Eb}{E_b}							
\safemath{\EbNo}{\frac{\Eb}{\No}}
\safemath{\EsNo}{\frac{\Es}{\No}}

\DeclareMathOperator{\CHop}{\ensuremath{\opH}} 
\safemath{\tvir}{\rndh_{\CHop}}				
\safemath{\tvtf}{\rndl_{\CHop}}				
\safemath{\spf}{\rnds_{\CHop}}				
\safemath{\bff}{H_{\CHop}}					

\safemath{\ircf}{r_{h}}						
\safemath{\tftvcf}{r_{s}}					
\safemath{\tfcf}{r_{l}}						
\safemath{\bfcf}{r_{H}}						

\safemath{\tcorr}{c_h}						
\safemath{\scf}{c_{s}}						
\safemath{\tfcorr}{c_{l}}					
\safemath{\fcorr}{c_{H}}						

\safemath{\mi}{I}							
\safemath{\capacity}{C}						

\safemath{\normal}{\mathcal{N}}			
\safemath{\jpg}{\mathcal{CN}}			
\safemath{\mchain}{\leftrightarrow}		

\safemath{\dB}{\,\mathrm{dB}}
\safemath{\dBm}{\,\mathrm{dBm}}
\safemath{\Hz}{\,\mathrm{Hz}}
\safemath{\kHz}{\,\mathrm{kHz}}
\safemath{\MHz}{\,\mathrm{MHz}}
\safemath{\GHz}{\,\mathrm{GHz}}
\safemath{\s}{\,\mathrm{s}}
\safemath{\ms}{\,\mathrm{ms}}
\safemath{\mus}{\,\mathrm{\text{\textmu}s}}
\safemath{\ns}{\,\mathrm{ns}}
\safemath{\ps}{\,\mathrm{ps}}
\safemath{\meter}{\,\mathrm{m}}
\safemath{\mm}{\,\mathrm{mm}}
\safemath{\cm}{\,\mathrm{cm}}
\safemath{\m}{\,\mathrm{m}}
\safemath{\W}{\,\mathrm{W}}
\safemath{\mW}{\, \mathrm{mW}}
\safemath{\J}{\,\mathrm{J}}
\safemath{\K}{\,\mathrm{K}}
\safemath{\bit}{\,\mathrm{bit}}
\safemath{\nat}{\,\mathrm{nat}}

\safemath{\define}{\triangleq}			

\safemath{\equivalent}{\sim}
\safemath{\distas}{\sim}					
\safemath{\sdiff}{\Delta}				

\safemath{\reals}{\mathbb{R}}
\safemath{\positivereals}{\reals_{+}}
\safemath{\integers}{\mathbb{Z}}
\safemath{\posint}{\integers_{+}}
\safemath{\naturals}{\mathbb{N}}
\safemath{\posnaturals}{\naturals_{+}}
\safemath{\complexset}{\mathbb{C}}
\safemath{\rationals}{\mathbb{Q}}

\newcommand*{\fancyrefapplabelprefix}{app}		
\newcommand*{\fancyrefthmlabelprefix}{thm}		
\newcommand*{\fancyreflemlabelprefix}{lem}		
\newcommand*{\fancyrefcorlabelprefix}{cor}		
\newcommand*{\fancyrefdeflabelprefix}{def}		
\newcommand*{\fancyrefproplabelprefix}{prop}		
\newcommand*{\fancyrefexmpllabelprefix}{exmpl}
\newcommand*{\fancyrefalglabelprefix}{alg}		
\newcommand*{\fancyreftbllabelprefix}{tbl}		

\frefformat{vario}{\fancyrefseclabelprefix}{Sec.~#1}
\frefformat{vario}{\fancyrefthmlabelprefix}{Thm.~#1}
\frefformat{vario}{\fancyreftbllabelprefix}{Tbl.~#1}
\frefformat{vario}{\fancyreflemlabelprefix}{Lem.~#1}
\frefformat{vario}{\fancyrefcorlabelprefix}{Corr.~#1}
\frefformat{vario}{\fancyrefdeflabelprefix}{Def.~#1}
\frefformat{vario}{\fancyreffiglabelprefix}{Fig.~#1}
\frefformat{vario}{\fancyrefapplabelprefix}{App.~#1}
\frefformat{vario}{\fancyrefeqlabelprefix}{(#1)}
\frefformat{vario}{\fancyrefproplabelprefix}{Prop.~#1}
\frefformat{vario}{\fancyrefexmpllabelprefix}{Ex.~#1}
\frefformat{vario}{\fancyrefalglabelprefix}{Alg.~#1}

\safemath{\dictab}{[\,\dicta\,\,\dictb\,]}

\safemath{\ysig}{\bmy}
\safemath{\ysighat}{\hat{\ysig}}
\safemath{\ysigdim}{M}
\safemath{\xsig}{\bmx}
\safemath{\xsigdim}{N}
\safemath{\nx}{n_x}
\safemath{\zsig}{\bmz}
\safemath{\zsigdim}{\ysigdim}
\safemath{\rsig}{\bmr}
\safemath{\Adict}{\bA}
\safemath{\Adicttilde}{\widetilde{\Adict}}
\safemath{\Adictdim}{\outputdim\times\xsigdim}
\safemath{\avec}{\bma}
\safemath{\avectilde}{\tilde{\avec}}
\safemath{\Bdict}{\bB}
\safemath{\Bdicttilde}{\widetilde{\Bdict}}
\safemath{\Cdict}{\bC}
\safemath{\cvec}{\bmc}
\safemath{\Ddict}{\bD}
\safemath{\Ddictdim}{\ysigdim\times\xsigdim}
\safemath{\dvec}{\bmd}
\safemath{\Ddicttilde}{\widetilde{\bD}}
\safemath{\Bonb}{\bB}
\safemath{\bvec}{\bmb}
\safemath{\Bonbdim}{\ysigdim\times\ysigdim}
\safemath{\noise}{\bmn}
\safemath{\noisedim}{\ysigim}
\safemath{\err}{\bme}
\safemath{\errdim}{\ysigdim}
\safemath{\errset}{\setE}
\safemath{\nerr}{n_e}
\safemath{\delop}{\bP_\errset}
\safemath{\delopc}{\bP_{{\errset}^c}}

\safemath{\cplxi}{\imath}
\safemath{\cplxj}{\jmath}

\safemath{\dict}{\matD}
\safemath{\inputdim}{N}		
\safemath{\outputdim}{M}		
\safemath{\sparsity}{S}	
\safemath{\inputdimA}{{N_a}}	
\safemath{\inputdimB}{{N_b}}	
\safemath{\elemA}{{n_a}}	
\safemath{\elemB}{{n_b}}	
\safemath{\resA}{\matR_a}	
\safemath{\resB}{\matR_b}	
\safemath{\subD}{\matS} 
\safemath{\subA}{\matS_a} 
\safemath{\subB}{\matS_b} 
\safemath{\dicta}{\matA} 	
\safemath{\dictb}{\matB} 	
\safemath{\hollowS}{H}
\safemath{\hollowA}{H_a}
\safemath{\hollowB}{H_b}
\safemath{\cross}{Z}
\safemath{\coh}{\mu_d}			
\safemath{\coha}{\mu_a}			
\safemath{\cohb}{\mu_b}			
\safemath{\mubs}{\nu}	
\safemath{\cohm}{\mu_m} 
\safemath{\dictset}{\setD}	
\safemath{\dictsetp}{\dictset(\coh,\coha,\cohb)}	
\safemath{\dictsetgen}{\dictset_\text{gen}}
\safemath{\dictsetgenp}{\dictsetgen(\coh)}
\safemath{\dictsetonb}{\dictset_\text{onb}}
\safemath{\dictsetonbp}{\dictsetonb(\coh)}

\safemath{\leftside}{U}
\safemath{\rightsideA}{R_a}
\safemath{\rightsideB}{R_b}

\safemath{\indexS}{\setI_S} 

\safemath{\na}{n_a}			
\safemath{\nb}{n_b}			
\safemath{\coeffa}{p_i}	
\safemath{\coeffb}{q_j}	
\safemath{\seta}{\setP}		
\safemath{\setb}{\setQ}     
\safemath{\setw}{\setW}	
\safemath{\setz}{\setZ}	
\safemath{\cola}{\veca}		
\safemath{\colb}{\vecb}		
\safemath{\cold}{\vecd}		
\safemath{\inputvec}{\vecx} 	
\safemath{\error}{\vece}	
\safemath{\noiseout}{\vecz} 	
\safemath{\inputvecel}{x}
\safemath{\inputveca}{\vecx_a}
\safemath{\inputvecb}{\vecx_b}
\safemath{\outputvec}{\vecy}	
\safemath{\lambdamin}{\lambda_{\mathrm{min}}}

\safemath{\elltwo}{\ell_2}
\safemath{\ellone}{\ell_1}
\safemath{\ellzero}{\ell_0}
\safemath{\ellinf}{\ell_\infty}
\safemath{\ellinftilde}{\ell_{\widetilde\infty}}
\safemath{\licard}{Z(\coh,\coha,\cohb)}
\safemath{\xsol}{\hat{x}}
\safemath{\xbord}{x_b}		
\safemath{\xstat}{x_s}		
\safemath{\xstatLone}{\tilde{x}_s}
\safemath{\order}{\mathcal{O}} 
\safemath{\scales}{\Theta} 
\safemath{\ones}{\mathbf{1}} 
\safemath{\zeroes}{\mathbf{0}} 
\safemath{\thlone}{\kappa(\coh,\cohb)} 
\safemath{\constoneA}{\delta} 
\safemath{\constoneB}{\epsilon} 
\safemath{\nlarge}{L}				   
\safemath{\sumlarge}{S_\nlarge}
\safemath{\maxlarger}{P_\nlarge}	   
\safemath{\Pzero}{\textrm{P0}}	
\safemath{\Pone}{\textrm{P1}}
\safemath{\vecfir}{\vecw}			 
\safemath{\vecsec}{\vecz}
\safemath{\elvecfir}{w}              
\safemath{\elvecsec}{z}				 
\safemath{\nlargefir}{n}
\safemath{\normout}{\gamma}
\safemath{\auxfun}{h}
\safemath{\supp}{\textrm{supp}}

\safemath{\indexa}{\ell}
\safemath{\indexb}{r}
\safemath{\indexc}{i}
\safemath{\indexd}{j}

\safemath{\project}{P}

\safemath{\firstslotset}{\setU_1}  
\safemath{\secondslotset}{\setU_2} 
\safemath{\randomset}{\setS} 

\safemath{\Tran}{\textnormal{T}}
\safemath{\Herm}{\textnormal{H}}

\newcommand*{\fancyreflstlabelprefix}{lst}
\fancyrefaddcaptions{english}{%
  \providecommand*{\freflstname}{Listing}%
}
\frefformat{vario}{\fancyreflstlabelprefix}{%
  \freflstname\fancyrefdefaultspacing#1#3%
}

\begin{document}
\title{On the Impact of Site-Specific Training \\ for a Real-World 5G NR System}

\author{\IEEEauthorblockN{Reinhard Wiesmayr$^{*,\text{1}}$, Nuri Berke Baytekin$^{*,\text{1}}$, Chris Dick$^\text{2}$, and Christoph Studer$^\text{1}$}\\[0.2cm]
    \em $^*$equal contribution; $^\textnormal{1}$ETH Zurich, $^\textnormal{2}$NVIDIA; e-mail: wiesmayr@iis.ee.ethz.ch%
    \thanks{The authors used OpenAI GPT 5.6 Sol during preparation of this manuscript. All outputs were independently evaluated and verified by the authors, who take full responsibility for the content of the manuscript.}
    \thanks{We thank Ant\'onio Maia Barros for his support during measurements.}
    \thanks{We acknowledge NVIDIA for their sponsorship of this research.}
    }

\maketitle

\begin{abstract}
Site-specific training can improve wireless receiver performance without increasing computational complexity. However, real-world results have so far focused on fully trainable neural receivers and single-layer transmissions. We study site-specific finetuning of three receiver architectures: fully trainable neural, model-driven neural, and model-based. We train and evaluate these receivers using new measurements from a standard-compliant 5G~NR testbed at ETH Zurich with dual-layer uplink transmission, including measurement campaigns conducted more than six months apart. Our results show that site-specific finetuning (i) substantially improves fully trainable and model-driven neural receivers, while resulting in only marginal gains for the less tunable model-based receiver; (ii) enables a single neural receiver jointly finetuned for single- and dual-layer transmission to closely match receivers finetuned separately for each configuration; and (iii) remains effective across measurement campaigns separated by more than six months. We also investigate site-specific linear minimum mean-square error channel estimation using covariance matrices estimated from either synthetic channels or site-specific measurements. When combined with iterative detection and decoding, site-specific channel estimation achieves the lowest error rate observed in our datasets. Our finetuning code and measurement datasets are publicly available online.\footnote{\label{fn:repository}\url{https://github.com/IIP-Group/site_specific_training}}
\end{abstract}

\glsresetall

\section{Introduction}

Site-specific training optimizes wireless receiver performance for a specific deployment area by exploiting the characteristic propagation conditions and \gls{RF} hardware impairments.
There exist two common approaches: (i)~\emph{site-specific finetuning}, where receiver algorithm parameters are tuned through \gls{SGD} using labeled datasets, and (ii)~\emph{site-specific parameter adaptation}, where  parameters of classical receiver algorithms are estimated for a given site.
Site-specific finetuning has been applied to several \gls{MIMO} data detectors:
fully-tunable \glspl{NRX}~\cite{honkala2021deeprx,honkala2021deeprxmimo,cammerer2023neuralreceiver5gnr}, which provide the highest tunability but also exhibit the highest complexity; model-based receivers~\cite{Wiesmayr2022}, which build upon low-complexity classical algorithms and introduce only a handful of hyperparameters tuned through deep unfolding~\cite{BalatsoukasStimming2019}; and \glspl{MDX}~\cite{abdollahpour2025modelnn}, which strike a balance between flexibility and complexity.
Site-specific parameter adaptation includes channel estimation algorithms that leverage channel statistics estimated for the given site~\cite{fesl2022channel}.

Most results on trainable receiver algorithms rely on synthetic data \cite{honkala2021deeprx,honkala2021deeprxmimo, cammerer2023neuralreceiver5gnr, fesl2022channel}.
However, reference~\cite{bock2025wireless} shows that synthetic-data-based results may be overly optimistic.
The study in~\cite{luostari2025adapting} trains an \gls{NRX} on synthetic data and validates its performance with real-world measurements from a custom single-antenna \gls{SDR} setup; this study reveals that \glspl{NRX} achieve better results when trained with channel properties that match the specific site.
Our initial experiments for site-specific \gls{NRX} finetuning in~\cite{baytekin2026site} show that finetuning with real-world measurement data is highly effective and halves the \gls{BLER} even on datasets from a \gls{UE} that was not used for training.
However, these initial results were limited to one receiver architecture and to single-user, single-layer \gls{MIMO} transmission.
    
\subsection{Contributions}

This paper investigates the following questions: \emph{Question~1:} Do the significant finetuning gains reported in~\cite{baytekin2026site}
also apply to (i)~other receiver architectures, (ii)~dual-layer transmission, and (iii)~temporally separated measurement campaigns? \emph{Question~2:} How do the gains from site-specific finetuning compare with those from site-specific parameter adaptation?
To answer Question~1, we assess the efficacy of site-specific finetuning across three receiver architectures: (i)~the fully-tunable \gls{NRX} from~\cite{cammerer2023neuralreceiver5gnr}, (ii)~the \gls{MDX} from~\cite{abdollahpour2025modelnn}, and (iii)~the model-based \gls{DUIDD} receiver from~\cite{Wiesmayr2022}.
To evaluate the efficacy of finetuning for single- and/or dual-layer transmission and its temporal robustness across measurement campaigns conducted more than half a year apart, we utilize existing as well as new \gls{OTA} measurement datasets from the standard-compliant \gls{5GNR} testbed at ETH Zurich~\cite{wiesmayr2025csi} with single- and dual-layer transmission data. 
To answer Question~2, we study site-specific parameter adaptation by comparing \gls{LMMSE} channel estimation with covariance matrices estimated from synthetic and site-specific measured data for non-iterative and iterative receivers.
The measurement datasets and finetuning code required to reproduce the results of this paper are available online.\textsuperscript{\ref{fn:repository}}

\subsection{Related Work}

Trainable receivers are typically evaluated on stochastic channel models~\cite{honkala2021deeprx,honkala2021deeprxmimo,cammerer2023neuralreceiver5gnr,abdollahpour2025modelnn,fesl2022channel,Wiesmayr2022}. Site-specific ray-tracing data has been used for \gls{NRX} finetuning in~\cite{wiesmayr2025design} and augmented with stochastic channel models~\cite{durfee2025sionnart}.
Finetuning results with real-world data are limited to our own initial experiments in~\cite{baytekin2026site} and subsequent independent work~\cite{luostari2026input}.
Our initial work~\cite{baytekin2026site} proposed a \gls{HARQ}-based data-extraction and \gls{NRX}-finetuning pipeline and demonstrated real-world gains across several deployment scenarios. The work in~\cite{luostari2026input} reports gains from tuning a different \gls{NRX} architecture with synthetic and measured data and bit labels only from successful transmissions on a custom single-antenna \gls{SDR} setup.
In contrast, we extend our initial results~\cite{baytekin2026site} to a range of different receiver architectures, dual-layer transmission, and temporally separated measurement campaigns.

Classical receiver algorithms can be adapted to a specific deployment site, e.g., by estimating site-specific statistical parameters; for example, \gls{LMMSE} channel estimation relies on channel covariance matrices~\cite{savaux2017lmmse}.
Data-driven channel estimators based on \glspl{GMM} also rely on channel statistics, but the approach in~\cite{fesl2022channel} was trained and evaluated on synthetic channel data only. Recent \gls{OTA} results show that gains from training a neural channel estimator with measured data are limited when noisy channel estimates are used as labels~\cite{luostari2026input}.
In contrast, we estimate site-specific covariance matrices from noisy \gls{LS} observations by properly accounting for the noise statistics, and we compare the resulting receiver performance with that obtained using covariance estimates from synthetic channels.

\section{Tunable Receiver Architectures}

We focus on three receiver architectures for our finetuning experiments that all utilize \gls{LS} channel estimates; more powerful \gls{LMMSE} channel estimators are studied in \fref{sec:covariance_based_receiver_adaptation}.

\subsection{Fully-Tunable Neural Receiver (NRX)}

The first receiver is the fully-tunable \gls{NRX} that implements the \gls{CGNN} architecture for \gls{5GNR} \gls{PUSCH} reception from \cite{cammerer2023neuralreceiver5gnr} combined with the multi-loss training scheme from \cite[Sec.~IV-B]{wiesmayr2025design}. In this architecture, a convolutional state-initialization layer processes the received \gls{OFDM} resource grid, initial \gls{LS} channel estimates, and a positional encoding that indicates the distance from each \gls{RE} to the next \gls{DMRS}. Consecutive unrolled \gls{CGNN} iterations combine message passing with convolutional state updates.
Readout layers transform the final state to soft-outputs for the coded bits.
We use the Deep \gls{NRX} with eight unrolled \gls{CGNN} iterations (denoted ``Large \gls{NRX}'' in \cite{wiesmayr2025design}) for initial training (denoted pretraining) and subsequent finetuning.
The pretrained and finetuned Deep \gls{NRX} model also yields the Shallow \gls{NRX} by stopping after two unrolled iterations during evaluation; the Shallow and Deep \gls{NRX} use approximately $1.2\cdot10^5$ and $4.4\cdot10^5$ active parameters, respectively.

\subsection{Model-Driven Neural Receiver (MDX)}

The second receiver is the \gls{MDX} from~\cite{abdollahpour2025modelnn} that combines conventional receiver algorithms with a small \gls{NN} to implement joint channel estimation and data detection. In this architecture, one first obtains pilot-aided \gls{LS} channel estimates from the \glspl{DMRS} and data-symbol estimates using conventional \gls{LMMSE} equalization. For each layer, the receiver subtracts the estimated interference from all other layers and reuses the data-symbol estimates as pilots to obtain data-aided (and hopefully improved) \gls{LS} channel estimates.
A two-dimensional \gls{ResNet} with four residual blocks then combines the pilot-aided and data-aided channel estimates. Finally, the resulting refined estimates are used for final \gls{LMMSE} max-log soft-output data detection~\cite{abdollahpour2025modelnn}. The tunable parameters comprise the weights of the \gls{ResNet} and coefficients that scale the noise variances used by the two \gls{LMMSE} equalizers and the soft-output stage; the evaluated implementation contains approximately $2.5\cdot10^3$ tunable parameters.

\subsection{Model-Based DUIDD Receiver}

The third receiver is model-based and derived from classical \gls{IDD}~\cite{HT2003}, which iteratively exchanges soft information between a soft-input soft-output MMSE-PIC detector~\cite{studer2011asic} and the \gls{LDPC} \gls{MP} decoder. The \gls{DUIDD} architecture interleaves data detection and channel decoding and applies deep unfolding to make the soft-information exchange, decoder message damping, and decoder state forwarding tunable~\cite{Wiesmayr2022}.
In our site-specific finetuning experiments, we evaluate untuned classical \gls{IDD} as a baseline and compare against pretrained and finetuned \gls{DUIDD} receivers, each of which uses $I=2$ MMSE-PIC data detection stages. We also evaluate classical \gls{IDD} with $I=1$, for which MMSE-PIC reduces to non-iterative \gls{LMMSE} detection.
Each MMSE-PIC detection stage is followed by $N_i=N_{\text{MP}}/I$ \gls{LDPC} \gls{MP} decoding iterations.
Hence, we fix the total number of \gls{MP} decoding iterations to $N_{\text{MP}}=12$ for all model-based \gls{IDD} and \gls{DUIDD} receivers investigated.
\gls{IDD} and \gls{DUIDD} receivers use flooding min-sum \gls{CN} updates implemented in NVIDIA Sionna; \gls{DUIDD} additionally applies tunable message damping from~\cite{Wiesmayr2022} and contains only 30 tunable parameters.
The \gls{DUIDD} receiver---in contrast to the \gls{NRX} and the \gls{MDX}---includes the \gls{LDPC} \gls{MP} decoder during tuning; the \gls{NRX} and the \gls{MDX} denote tunable \gls{MIMO} detectors that are combined with an external decoder for evaluation. For the \gls{NRX}, the \gls{MDX}, and the powerful real-time NVIDIA Aerial receiver~\cite{pusch_channel_estimator} (denoted ``MMSE Reference Rx''), we utilize the \gls{LDPC} \gls{MP} decoder from NVIDIA pyAerial with $N_{\text{MP}}=10$ iterations using layered normalized min-sum \gls{CN} updates.\footnote{\gls{LDPC} \gls{MP} decoding with 10 layered iterations usually performs better than 12 flooding iterations. However, this choice was an implementation constraint.}

\section{Site-Specific Deployment Scenarios}\label{sec:measurements}

\begin{figure}[tp]
\centering
\begin{subfigure}[t]{\columnwidth}
    \centering
    \includegraphics[height=4cm]{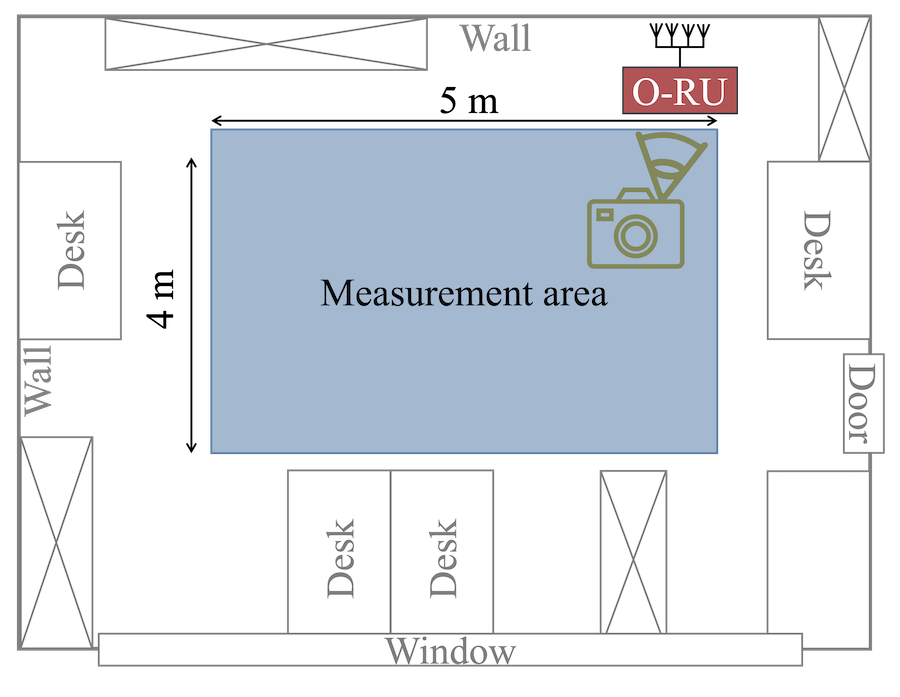}\hfill
    \includegraphics[height=4cm]{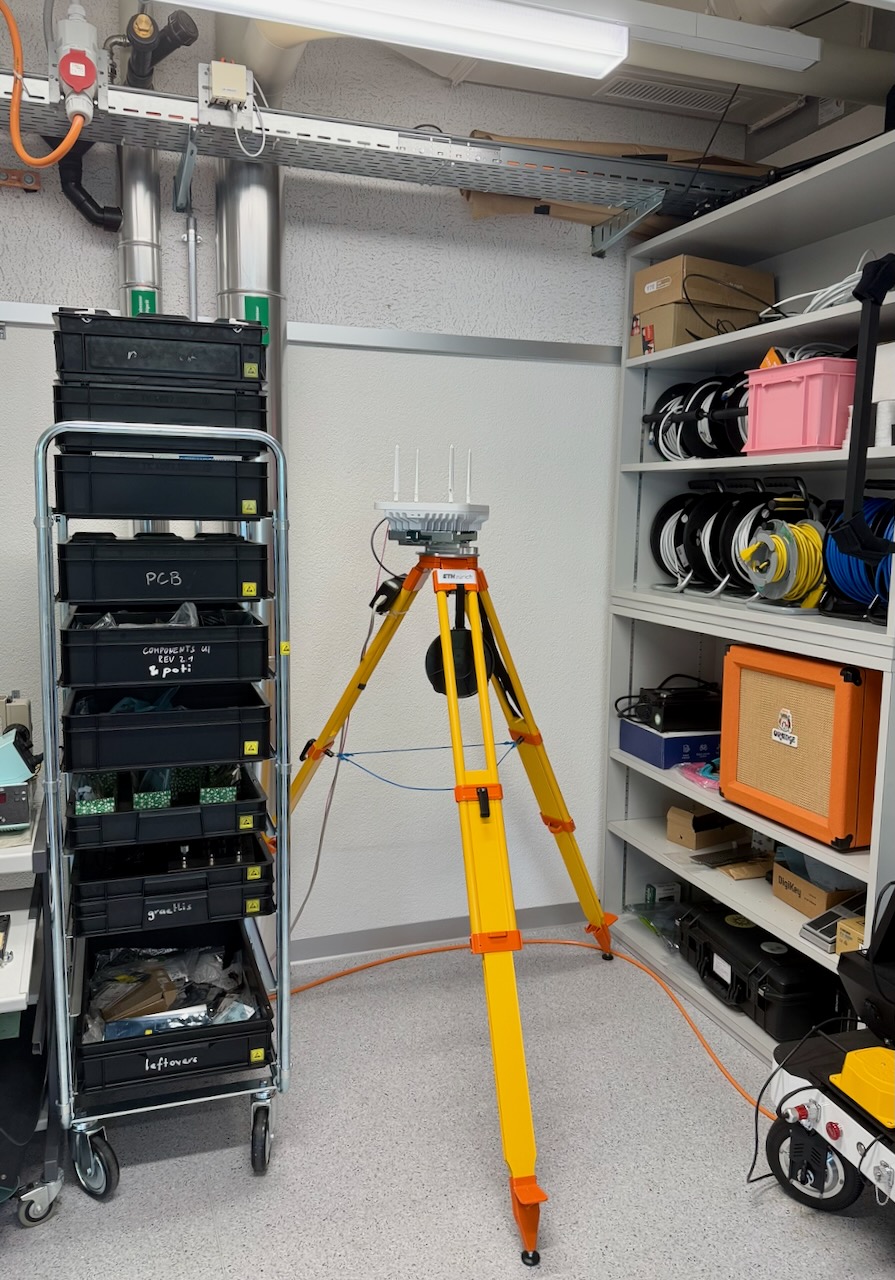}
    \caption{Indoor small laboratory}
    \label{fig:deployment_scenarios_small_lab}
\end{subfigure}\\[0.2cm]
\begin{subfigure}[t]{\columnwidth}
    \centering
    \includegraphics[height=3.75cm]{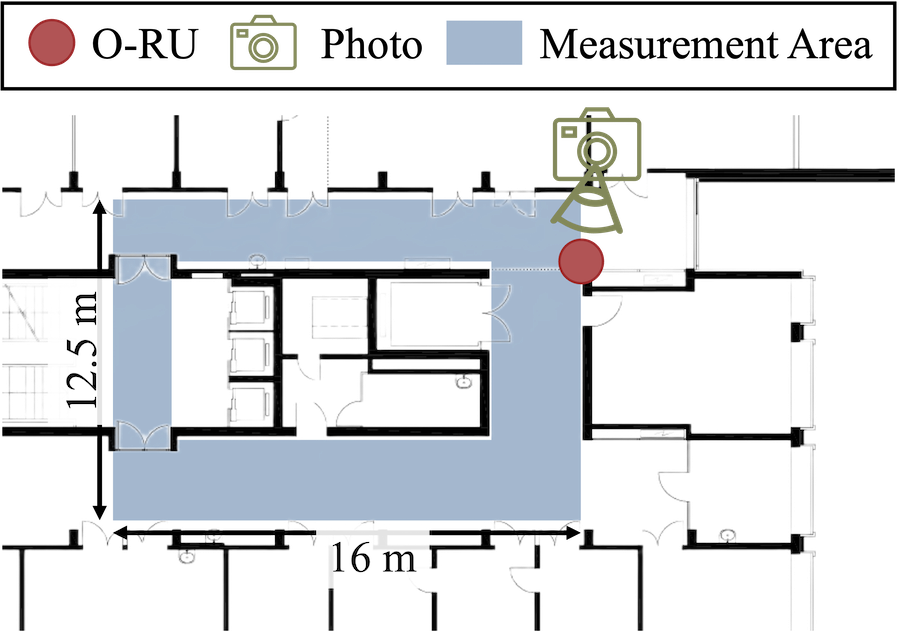}\hfill
    \includegraphics[height=3.75cm]{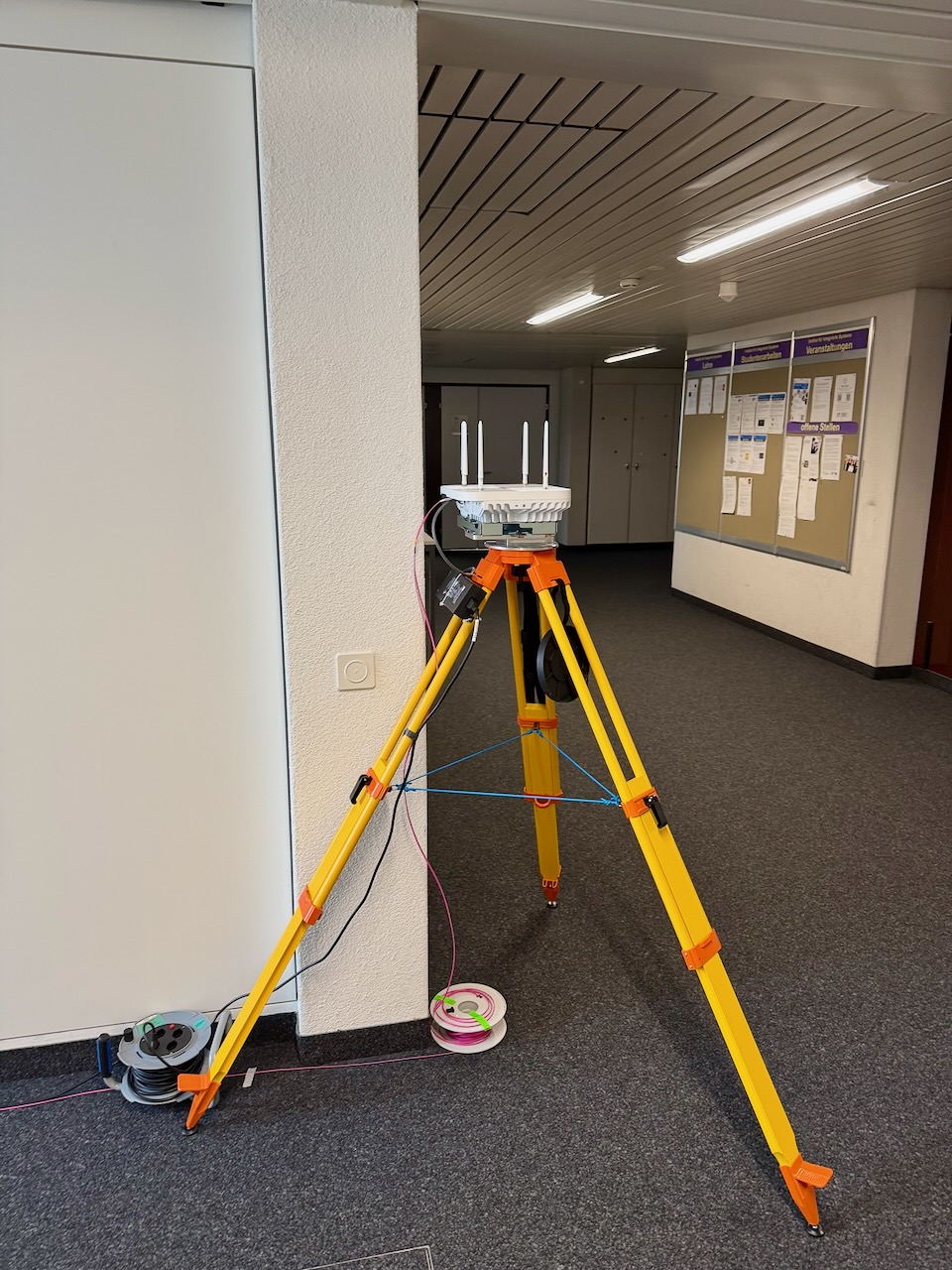}
    \caption{Indoor large office floor}
    \label{fig:deployment_scenarios_institute_floor}
\end{subfigure}
\caption{Floor plans (left) and photographs of the serving \glspl{ORU} (right) for the two indoor deployment scenarios in June 2026.}
\label{fig:deployment_scenarios}
\end{figure}

We now describe the various deployment scenarios considered in this paper. 
All datasets consist of \emph{\gls{PUSCH} slot samples}, each comprising the \gls{OFDM}-domain receive signals of one complete \gls{UL} transmission slot. We obtain the measurements with \gls{COTS} \glspl{UE} on the standard-compliant \gls{5GNR} testbed at ETH Zurich~\cite{wiesmayr2025csi} in the two indoor deployment scenarios shown in \fref{fig:deployment_scenarios}. The testbed builds upon the NVIDIA Aerial Testbed (ATB) software platform~\cite{nvidia_ATB_2025} and uses \gls{COTS} \glspl{ORU} with four antennas each, 100\,MHz bandwidth centered at 3.45\,GHz with 273 \glspl{PRB}, 30\,kHz subcarrier spacing, and the \gls{TDD} slot pattern 3DSU (3\,\gls{DL}, 1 special, 1\,\gls{UL}). The \texttt{iperf3} tool was used to generate continuous application-layer uplink traffic while carrying one \gls{UE} at a time along a random trajectory through the measurement area. The testbed records each slot sample from the serving \gls{ORU}'s fronthaul.

\begin{figure}[tp]
    \centering
    \includegraphics[width=.8\linewidth]{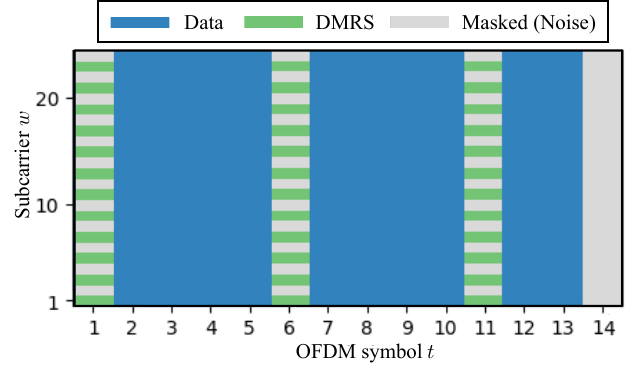}
    \caption{\gls{OFDM} resource grid of the slot samples used for finetuning and testing, showing 24 of 3276 subcarriers and \glspl{DMRS} from the first layer.
    \glspl{DMRS} from the second layer occupy the even subcarriers of \gls{OFDM} symbols $t\in\{1,6,11\}$.}
    \label{fig:resource_grid}
\end{figure}

The slot samples used for finetuning and evaluation follow the \gls{OFDM} resource grid shown in \fref{fig:resource_grid}.
The testbed is configured for low \gls{PUSCH} \gls{SNR} targets for which the link-adaptation algorithm, configured for a target \gls{BLER} of 10\,\%, schedules 16-QAM transmissions. Hence, our measurements capture challenging operating points in the MMSE Reference Rx's \gls{BLER}-versus-\gls{SNR} waterfall region.

\subsection{Indoor Small Laboratory Scenario}

\subsubsection{Dataset from Nov.\ 2025 with Single-Layer Transmission}

The first measurement campaign was conducted in the indoor small laboratory in November 2025 and was previously described in~\cite{baytekin2026site}. This scenario covers a measurement area of about $3.5\,\text{m}\times3.5\,\text{m}$ that is surrounded by desks and laboratory equipment. Four \glspl{ORU} are placed around the measurement area; the serving \gls{ORU} is located in the upper-right corner, as in the similar setup shown in~\fref{fig:deployment_scenarios_small_lab}. The measurement campaign comprises two single-layer measurements, one with a Samsung Galaxy S23 and one with an Apple iPhone 14 Pro. The \gls{PUSCH} \gls{SNR} target is 7\,dB.

\subsubsection{Dataset from Jun.\ 2026 with Single- and Dual-Layer Transmission}

The second measurement campaign was conducted in June 2026, again in the same small-laboratory environment, more than half a year after the campaign discussed above. The scenario now covers a $5\,\text{m}\times4\,\text{m}$ measurement area with only a single serving \gls{ORU} and no passive listeners.
This campaign comprises two measurements. The single-layer measurement uses a Samsung Galaxy S23 with a 5\,dB \gls{PUSCH} \gls{SNR} target; the dual-layer measurement uses a Google Pixel 9 Pro with a 12\,dB target. Single-layer transmissions always use \gls{DMRS} port~0 in \gls{CDM} group~0, whereas dual-layer transmissions use \gls{DMRS} ports~0 and~2 in \gls{CDM} groups~0 and~1, respectively.

\subsection{Indoor Large Office Floor Scenario}

\subsubsection{Dataset from Nov.\ 2025 with Single-Layer Transmission}

The first measurement campaign was conducted on the indoor large office floor in November 2025 and was previously described in~\cite{baytekin2026site}. This scenario is about $50\,\text{m}\times30\,\text{m}$ and offers a measurement area that covers the corridors and some adjacent rooms. The scenario provides both \gls{LOS} and \gls{NLOS} propagation conditions. The campaign comprises two single-layer measurements with the same two \glspl{UE} and \gls{SNR} targets as their November 2025 small-laboratory counterparts.

\subsubsection{Dataset from Jun.\ 2026 with Single- and Dual-Layer Transmission}

The second measurement campaign was conducted in June 2026, again in the same large-office-floor environment, more than half a year after the November 2025 campaign discussed above. The scenario now covers a smaller section of the office floor, comprising a corridor area of approximately $16\,\text{m}\times12.5\,\text{m}$ with the serving \gls{ORU} deployment shown in \fref{fig:deployment_scenarios_institute_floor}. The single- and dual-layer measurements use the same \glspl{UE}, \gls{ULL} configurations, \gls{SNR} targets, and \gls{DMRS} configuration as their June 2026 small-laboratory counterparts.

\section{Evaluation Protocol}

We now detail the data extraction procedure, the site-specific finetuning procedure, and our performance evaluation approach. 

\subsection{Finetuning Data Extraction Pipeline}

We extend the site-specific finetuning and evaluation pipeline from~\cite{baytekin2026site} to the receiver architectures and \gls{ULL} configurations considered here.
For every recorded slot sample, the pipeline extracts the \gls{OFDM}-domain receive signals and the corresponding scheduling and receiver configuration from NVIDIA Data Lake~\cite{nvidia_aerial_datalake}. Ground-truth bit labels for successful transmissions are obtained from the decoded payload bits. For failed transmissions, we track their \gls{HARQ} processes and obtain the payload bits from later successful retransmissions. We then re-encode the payload bits using the redundancy version and scrambling configuration of the initial transmission to reconstruct its coded bit labels~\cite{baytekin2026site}. We retain slot samples with 16-QAM and \gls{MCS} indices 6--10 from~\cite[Table 5.1.3.1-2]{38214}.
For \gls{DUIDD} finetuning and the corresponding receiver comparison in \fref{fig:results-duidd-ri2}, we restrict the data to \gls{MCS} index~10 and one scrambling configuration. The \gls{MDX}, the non-iterative \gls{LMMSE} receiver, and the \gls{IDD} and \gls{DUIDD} receivers are provided with a scalar pre-equalization wideband noise-variance estimate returned by the NVIDIA pyAerial noise-and-interference estimator~\cite{nvidia_pyaerial}; the \gls{NRX} does not need such a noise-variance estimate.

\subsection{Dataset Splits for Finetuning and Testing}\label{sec:dataset_splits}

For each measurement campaign and \gls{ULL} configuration, we form disjoint datasets for finetuning and testing. The November 2025 datasets from our previous work~\cite{baytekin2026site} use cross-\gls{UE} finetuning and test data splits; each new June 2026 dataset is split at a manually selected timestamp into two temporally disjoint finetuning and test datasets, because each June 2026 measurement uses only one \gls{UE}. We construct each finetuning and test dataset such that approximately 10\,\% of its transport blocks\footnote{A transport block comprises all coded bits transmitted within a slot.} fail with the MMSE Reference Rx. All receivers within a comparison are evaluated on the same test slot samples.

\subsection{Site-Specific Finetuning Pipeline}

We initialize the finetuning pipeline with receiver parameters pretrained as follows.
We pretrain the \gls{NRX} on randomized 3GPP UMi channels using the single- and dual-user configurations from~\cite{baytekin2026site,wiesmayr2025design} and adapt its frame structure to match the resource grid in \fref{fig:resource_grid}. We pretrain the \gls{MDX} on synthetic 3GPP UMi channels using a single-user, single-layer transmission configuration and the same frame structure. We pretrain the \gls{DUIDD} receiver on synthetic 3GPP UMi channels and dual-layer transmission for $6\cdot10^3$ batches: $3\cdot10^3$ batches with the \gls{BCE} loss followed by $3\cdot10^3$ batches with the normalized LogSumExp \gls{BLER} loss~\cite{Wiesmayr2023}.

After pretraining, we finetune the \gls{NRX} and \gls{MDX} with the \gls{BCE} loss for $10^5$ batches of 16 slot samples spanning four consecutive \glspl{PRB}, which are
cropped from the full-band slot samples with 273~\glspl{PRB}.
We finetune the \gls{DUIDD} receiver for $2\cdot10^3$ batches of 16 full-band slot samples: the first $10^3$ batches use the \gls{BCE} loss, and the subsequent $10^3$ batches use the normalized LogSumExp \gls{BLER} loss.

\subsection{Site-Specific Performance Evaluation}\label{sec:perf_eval}

For site-specific performance evaluation, we first study the \emph{dataset \gls{BLER}} at the measured \gls{SNR} defined by the \gls{PUSCH} \gls{SNR} target~\cite[Sec.~V-C]{baytekin2026site}.
The \emph{dataset \gls{BLER}} is the fraction of transport blocks that fail after offline receiver processing and \gls{LDPC} decoding on a fixed test dataset. Remember that the test dataset is itself sampled from measurements such that the MMSE Reference Rx achieves a dataset \gls{BLER} of about 10\,\% (cf.\ \fref{sec:dataset_splits}).
In contrast to system-level \gls{BLER} under link adaptation, the dataset \gls{BLER} enables fair receiver comparisons on the same recorded transmissions.

Second, we study the dataset \gls{BLER} over a range of \emph{effective \gls{SNR}} values by adding complex white Gaussian noise to the test dataset~\cite[Sec.~V-C]{baytekin2026site}.
The recorded \gls{OFDM}-domain received signal follows the model $y=s+n$, where $s\in\complexset$ is the noise-free signal component from all active \glspl{ULL} with aggregate power $E_s$ and $n\in\complexset$ is the receiver noise with variance $N_0$.
We add complex Gaussian noise $z\sim\jpg(0,N_z)$ according to $\tilde y=y+z$, where $N_z=\alpha(E_s+N_0)$ and $\alpha\geq0$. The tested receiver, including its \gls{LS} channel estimator, then processes~$\tilde y$. For the original effective \gls{SNR} $\gamma=E_s/N_0$ (cf. the \gls{PUSCH} \gls{SNR} targets defined in \fref{sec:measurements}), the resulting effective \gls{SNR} is
\begin{equation}
    \text{SNR}_{\text{eff}}
    =\frac{E_s}{N_0+\alpha(E_s+N_0)}
    =\frac{\gamma}{1+\alpha(\gamma+1)}.\label{eq:effective_snr}
\end{equation}
Sweeping $\alpha\geq0$ decreases the effective \gls{SNR} and yields dataset \gls{BLER} versus effective \gls{SNR} curves.

\section{Receiver Finetuning Results}\label{sec:receiver_finetuning_results}

The \gls{NRX}, \gls{MDX}, non-iterative \gls{LMMSE}, \gls{IDD}, and \gls{DUIDD} receivers evaluated in this section all use \gls{LS} channel estimates obtained with the standard \gls{PUSCH} \gls{LS} channel estimator from Sionna v0.19.2. This estimator applies frequency-domain orthogonal cover-code (OCC) combining by averaging each pair of channel estimates from adjacent pilot subcarriers; we refer to its outputs as post-OCC-combined \gls{LS} estimates. The MMSE Reference Rx instead uses multi-stage \gls{MMSE} channel estimation with delay estimation~\cite{pusch_channel_estimator}.

\subsection{Results With Single-Layer Transmission}

\subsubsection{Model-Driven Neural Receiver Finetuning}\label{sec:mdx_results}

\begin{figure}[tp]
    \centering
    \includegraphics[width=0.99\linewidth]{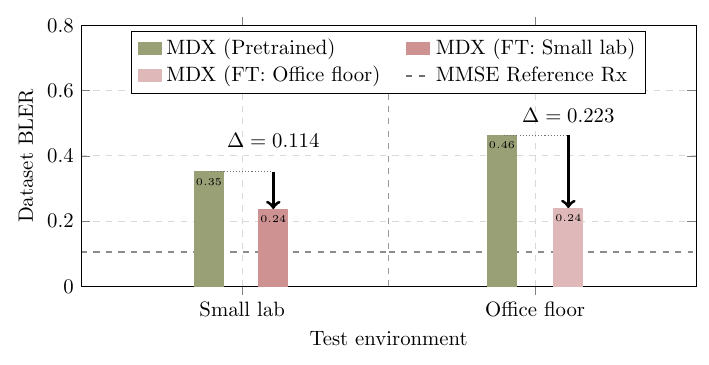}
    \caption{Dataset \gls{BLER} of the pretrained and finetuned \gls{MDX} on the small laboratory (left) and office floor (right) measurements recorded in June 2026 with single-layer transmission.}
    \label{fig:results-mdx-ri1}
\end{figure}

We start by analyzing the impact of site-specific finetuning on the \gls{MDX}. \fref{fig:results-mdx-ri1} compares the pretrained \gls{MDX} with a separate \gls{MDX} finetuned for each of the single-layer small-laboratory and office-floor scenarios from June 2026. Site-specific finetuning reduces the dataset \gls{BLER} by about one third (from 0.351 to 0.237) in the small laboratory and by nearly one half (from 0.463 to 0.240) on the office floor. The MMSE Reference Rx achieves a lower dataset \gls{BLER} than the finetuned \gls{MDX} in both environments. On the same small-laboratory test data from June 2026, \fref{fig:nrx-temporal-robustness-of-finetuning} shows a similar relative reduction by about one third from \gls{NRX} finetuning. Both the pretrained and finetuned \glspl{NRX} nevertheless achieve lower absolute dataset \gls{BLER} than their \gls{MDX} counterparts, at the cost of substantially more tunable parameters and higher inference latency~\cite{abdollahpour2025modelnn}.

\subsubsection{Temporal Robustness of NRX Finetuning}\label{sec:temporal_robustness}

\begin{figure}[tp]
    \centering
    
    \begin{subfigure}{0.99\linewidth}
        \centering
        \includegraphics[width=\linewidth]{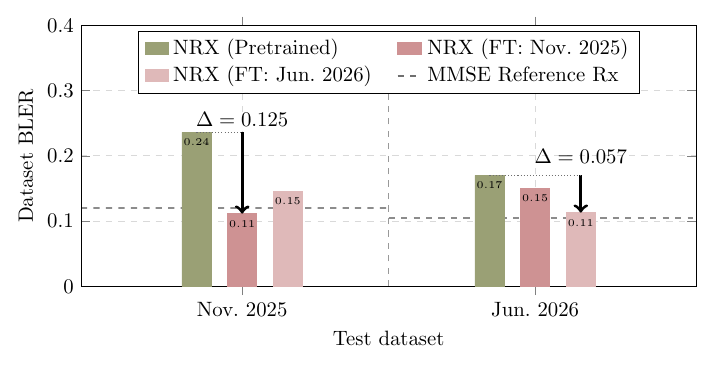}
        \vspace{-0.6cm}
        \caption{Shallow NRX (2 iter.)}
        \label{fig:results-nrx-temporal-shallow}
    \end{subfigure}
    \begin{subfigure}{0.99\linewidth}
        \centering
        \includegraphics[width=\linewidth]{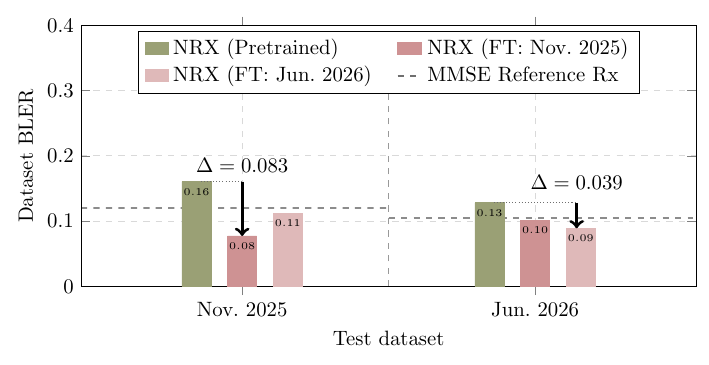}
        \vspace{-0.6cm}
        \caption{Deep NRX (8 iter.)}
        \label{fig:results-nrx-temporal-deep}
    \end{subfigure}
    \caption{Dataset \gls{BLER} of the pretrained and finetuned Shallow \gls{NRX} and Deep \gls{NRX} on the small laboratory datasets from November 2025 (left) and June 2026 (right) with single-layer transmissions.
    The legend identifies the finetuning dataset for each \gls{NRX} (``FT: Nov.\ 2025'' or ``FT: Jun.\ 2026'').}
    \label{fig:nrx-temporal-robustness-of-finetuning}
\end{figure}

We now investigate the temporal robustness of site-specific \gls{NRX} finetuning using data from measurement campaigns conducted more than half a year apart. \fref{fig:nrx-temporal-robustness-of-finetuning} compares the pretrained and finetuned Shallow \gls{NRX} (top) and Deep \gls{NRX} (bottom) on the single-layer small-laboratory data from November 2025 (left) and June 2026 (right). A separate model is finetuned on each measurement campaign and evaluated on both campaigns. For both \gls{NRX} configurations, using data from the same campaign for finetuning and testing yields the lowest dataset \gls{BLER}.
A temporal gap between measurement campaigns for finetuning and testing increases the dataset \gls{BLER}, most visibly for the Shallow \gls{NRX} tested on the June 2026 data; nevertheless, an \gls{NRX} finetuned on either campaign outperforms the pretrained \gls{NRX} when tested on the other campaign. These observations demonstrate that site-specific finetuning remains effective over time but motivate continued finetuning after deployment.

\subsection{Results With Dual-Layer Transmission}

\subsubsection{Finetuning on Dual-Layer vs. Single- and Dual-Layer Data}

\begin{figure}[tp]
    \centering
    \includegraphics[width=0.99\linewidth]{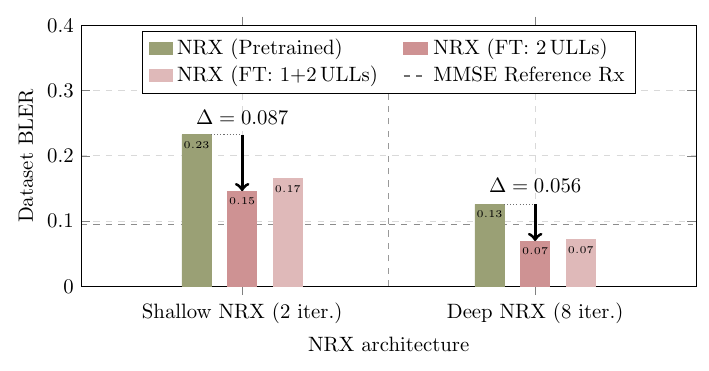}
    \caption{Dataset \gls{BLER} of the pretrained and finetuned Shallow \gls{NRX} (2~iter.) and Deep \gls{NRX} (8~iter.) for dual-layer transmission on the small laboratory dataset from June 2026.
    The \glspl{NRX} are finetuned either on dual-layer data (``FT:~2\,ULLs'') or on both single- and dual-layer data (``FT:~1+2\,ULLs'').}
    \label{fig:results-nrx-ri2}
\end{figure}

We now investigate in \fref{fig:results-nrx-ri2} the effectiveness of \gls{NRX} finetuning for single- and/or dual-layer transmission, evaluated on dual-layer transmission. Finetuning only on dual-layer data reduces the dataset \gls{BLER} by more than one third for both \gls{NRX} depths. The finetuned Deep \gls{NRX} clearly outperforms the MMSE Reference Rx, whereas the finetuned Shallow \gls{NRX} still has a higher dataset \gls{BLER}.
Finetuning on both single- and dual-layer data yields dataset \glspl{BLER} close to those obtained by finetuning only on dual-layer data with an absolute difference of at most 0.019.
The difference is negligible for the Deep \gls{NRX} and slightly larger for the Shallow \gls{NRX}.

A further experiment not presented in this paper confirms a similar effectiveness of \gls{NRX} finetuning for single- and/or dual-layer transmission, when evaluated on single-layer transmission. 
Finetuning on both single- and dual-layer data yields dataset \glspl{BLER} close to those obtained by finetuning only on single-layer data with an absolute difference of at most 0.004.

\subsubsection{Effective SNR Improvement of NRX Finetuning}

\begin{figure}[tp]
    \centering
    \includegraphics[width=0.99\linewidth]{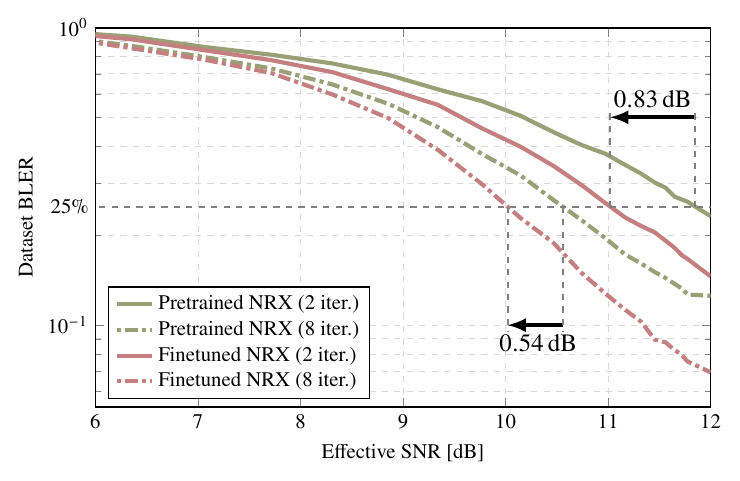}
    \caption{Dataset \gls{BLER} performance of the pretrained and finetuned Shallow \gls{NRX} (2~iter.) and Deep \gls{NRX} (8~iter.) vs.\ effective \gls{SNR} on the small laboratory dataset from June 2026 with dual-layer transmission. The arrows indicate the effective \gls{SNR} improvements to achieve a dataset \gls{BLER} of 25\,\%.}
    \label{fig:bler-snr-ri2}
\end{figure}

We now investigate the effective \gls{SNR} improvements from site-specific \gls{NRX} finetuning under dual-layer transmission. \fref{fig:bler-snr-ri2} evaluates the pretrained and finetuned \glspl{NRX} over a range of effective \gls{SNR} values obtained by adding white Gaussian noise to the dual-layer small-laboratory dataset as described in \fref{sec:perf_eval}.
Finetuning improves the dataset \gls{BLER} throughout the evaluated waterfall region and reduces the required effective \gls{SNR} to achieve a dataset \gls{BLER} of 25\,\% by 0.83\,dB for the Shallow \gls{NRX} and by 0.54\,dB for the Deep \gls{NRX}.

\subsubsection{Model-Based DUIDD Receiver Finetuning}

\begin{figure}[tp]
    \centering
    \includegraphics[width=0.99\linewidth]{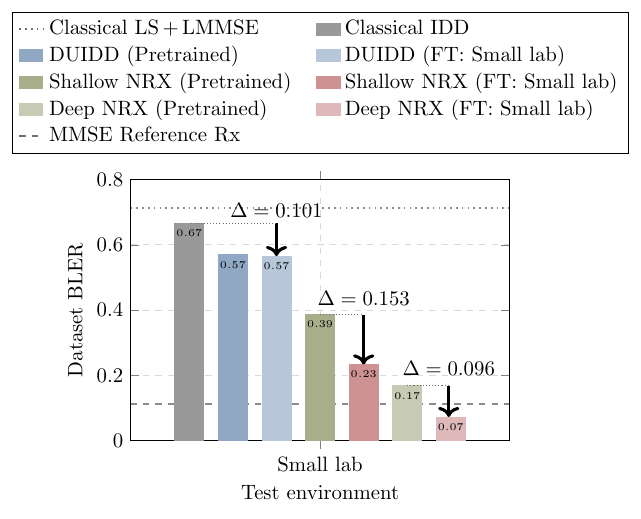}
    \caption{Dataset \gls{BLER} performance on the small laboratory dataset from June 2026 with dual-layer transmission and \gls{MCS} index~10. The \gls{IDD} and \gls{DUIDD} receivers use $I=2$ detection stages.}
    \label{fig:results-duidd-ri2}
\end{figure}

\fref{fig:results-duidd-ri2} compares the non-iterative \gls{LMMSE} receiver ($I=1$) with untuned classical \gls{IDD} and pretrained and finetuned \gls{DUIDD} receivers (all with $I=2$), as well as the Shallow and Deep \glspl{NRX} and the MMSE Reference Rx. All receivers are evaluated on the dual-layer small-laboratory data from June 2026, restricted to \gls{MCS} index~10. This is a different test subset from that used in \fref{fig:results-nrx-ri2}; the \gls{NRX} results are therefore not directly comparable.

Classical \gls{IDD} reduces the dataset \gls{BLER} compared with the non-iterative \gls{LMMSE} receiver, and the pretrained \gls{DUIDD} provides a further reduction. Site-specific finetuning of \gls{DUIDD}, however, reduces the dataset \gls{BLER} by only another 0.004 in absolute terms. This marginal gain suggests that the receiver's limited tunability leaves little room for further finetuning.

Both \gls{NRX} configurations achieve lower dataset \gls{BLER} and substantially larger finetuning gains than \gls{DUIDD}, but at much higher complexity. The \gls{DUIDD} receiver has similar complexity to classical \gls{IDD} and about twice the detection complexity of the non-iterative baseline~\cite[Sec.~V-D]{Wiesmayr2022}. The figure compares complete receiver chains: the \glspl{NRX} perform data detection with implicit channel refinement and use the NVIDIA pyAerial \gls{LDPC} \gls{MP} decoder with layered normalized min-sum \gls{CN} updates, whereas \gls{IDD} and \gls{DUIDD} operate on \gls{LS} channel estimates and use flooding min-sum \gls{CN} updates. The results in \fref{sec:covariance_based_receiver_adaptation} show substantial improvements when the model-based receivers instead use \gls{LMMSE} channel estimates, suggesting that better channel estimation has a greater impact on the dataset \gls{BLER} than improving the \gls{MIMO} detectors.

\section{Site-Specific Parameter Adaptation}\label{sec:covariance_based_receiver_adaptation}

We now study the impact of \gls{LMMSE} channel estimation on the dataset \gls{BLER} using covariance matrices estimated either from synthetic channels or from site-specific measurements.

\subsection{LMMSE Channel Estimation}

In contrast to \fref{sec:receiver_finetuning_results}, in which we focused on \gls{LS} channel estimation, we now use the \gls{LMMSE} interpolator from Sionna v0.19.2~\cite{Hoydis2022} to smoothen (i.e., denoise and interpolate) the \gls{LS} channel estimates. Sionna approximates the joint three-dimensional \gls{LMMSE} estimation problem by successive one-dimensional smoothing across the receive antennas, subcarriers, and \gls{OFDM} symbols, which we execute in this order.

For the considered \gls{DMRS} configurations, only one port is active in each \gls{CDM} group. Hence, every pilot provides a separate channel observation at its actual subcarrier. We modify Sionna's \gls{PUSCH} \gls{LS} channel estimator to omit orthogonal cover-code averaging and refer to its outputs as \emph{direct \gls{LS} estimates}. For one receive antenna, \gls{DMRS} symbol, and transmission layer, let $y[w]\in\complexset$ be the received \gls{OFDM}-domain signal on subcarrier $w$ and $p[w]\in\complexset$ the transmitted \gls{DMRS} symbol, including its orthogonal cover-code weight. The direct \gls{LS} estimate is given by
\begin{equation}
    \hat z[w]
    =\frac{y[w]}{p[w]}
    =h[w]+e[w],
    \label{eq:direct_ls_channel_estimate}
\end{equation}
where $h[w]\in\complexset$ is the channel coefficient and $e[w]\in\complexset$ the \gls{LS} estimation error.

For frequency-domain smoothing, let $\vech\in\complexset^W$ contain the channel coefficients over the $W=3276$ active subcarriers for one layer and receive antenna. Stacking the direct estimates from the $N_{\textnormal{p}}=1638$ pilot subcarriers of one \gls{DMRS} yields
\begin{align}\label{eq:observation_model}
    \hat\vecz=\matS\vech+\vece,
\end{align}
where $\matS\in\{0,1\}^{N_{\textnormal{p}}\times W}$ selects the pilot subcarriers with one nonzero entry per row, and $\vece\in\complexset^{N_{\textnormal{p}}}$ is the \gls{LS} estimation error. With the frequency-domain channel covariance $\matR_{\textnormal{f}}=\Ex{}{\vech\herm{\vech}}\in\complexset^{W\times W}$ and error covariance $\matSigma_{\textnormal{f}}\in\reals^{N_{\textnormal{p}}\times N_{\textnormal{p}}}$, the frequency-domain \gls{LMMSE} estimate is
\begin{equation}
    \hat{\vech}
    =\matR_{\textnormal{f}}\herm{\matS}
    \left(\matS\matR_{\textnormal{f}}\herm{\matS}+\matSigma_{\textnormal{f}}\right)^{-1}\hat\vecz.
    \label{eq:lmmse_channel_interpolation}
\end{equation}

Spatial and temporal smoothing work analogously and use the channel covariance matrices ${\matR}_{\textnormal{s}}\in\complexset^{4\times4}$ and ${\matR}_{\textnormal{t}}\in\complexset^{13\times13}$, respectively. We estimate the three channel covariance matrices for a given deployment site; the corresponding error covariance matrices $\matSigma_{\textnormal{s}}$, $\matSigma_{\textnormal{f}}$, and $\matSigma_{\textnormal{t}}$ are derived during runtime from the scalar pre-equalization wideband noise-variance estimate.

\subsection{Site-Specific Covariance Matrix Estimation}

As a baseline, we estimate the three covariance matrices from synthetic 3GPP UMi channel realizations~\cite{38901} using the implementation released with~\cite{wiesmayr2025design}. This implementation samples ground-truth full-band \gls{OFDM}-domain channels for randomized UMi topologies and averages their outer products over channel realizations and all remaining slot sample tensor dimensions not represented by each covariance matrix.

For site-specific estimation, we instead use the finetuning data with the resource-grid configuration in \fref{fig:resource_grid}. All three estimators follow the same principle: We form second moments from the direct \gls{LS} estimates in \eqref{eq:direct_ls_channel_estimate} and remove their estimation-error contribution. We estimate the required error statistics from the noise-only fourteenth \gls{OFDM} symbol after dividing its observations by the \gls{DMRS} sequence values used to obtain the corresponding direct \gls{LS} estimates. The dimension-specific estimation procedures and noise corrections are described next.

\subsubsection{Spatial Covariance Matrix}

For each slot sample, transmission layer, \gls{DMRS} symbol, and pilot subcarrier, we stack the direct \gls{LS} estimates from the four receive antennas into a vector.
Averaging the outer products of these vectors yields the spatial second moment. We form noise vectors analogously and average their outer products to obtain the spatial error covariance $\hat\matSigma_{\textnormal{s}}\in\complexset^{4\times4}$. Subtracting $\hat\matSigma_{\textnormal{s}}$ from the spatial second moment yields the estimated covariance matrix~$\hat{\matR}_{\textnormal{s}}$.

\subsubsection{Frequency Covariance Matrix}

The \glspl{DMRS} of each layer occupy only every second subcarrier. Hence, direct \gls{LS} estimates are unavailable on the remaining subcarriers, and we cannot estimate their cross-correlations directly. We therefore assume that all subcarriers share the same marginal frequency statistics and restrict $\hat{\matR}_{\textnormal{f}}$ to Hermitian Toeplitz form. This common simplification incurs only minor performance loss~\cite{fesl2022channel} and can be justified under the \gls{WSSUS} assumption~\cite{bock2024statistical}. Although we do not expect our channels to be \gls{WSSUS}, this restriction reduces the number of parameters from $W^2$ to $W$. We therefore estimate the correlation sequence $r[q]=\Ex{}{h[w+q]\conj{h}[w]}$ for $q\in\{-W+1,\ldots,W-1\}$.

We first estimate the available even lags. Let $\hat c_z[2q]$ denote the empirical correlation obtained by averaging $\hat z[w+2q]\conj{\hat z}[w]$ over all valid subcarrier indices, slot samples, receive antennas, transmission layers, and \gls{DMRS} symbols. We obtain the error correlation $\hat c_e[2q]$ by averaging $e[w+2q]\conj{e}[w]$ from the processed noise observations in the same manner. The noise-corrected even lags are
\begin{equation}
    \hat r[2q]
    =\hat c_z[2q]-\hat c_e[2q],
    \quad q=0,\ldots,N_{\textnormal{p}}-1.
    \label{eq:site_specific_even_lags}
\end{equation}

Since the effective channel impulse response is shorter than the \gls{CP}, its frequency response varies smoothly across subcarriers. We assume the same for its correlation sequence and obtain the missing odd lags by averaging neighboring even-lag estimates; for the final odd lag, we use the preceding even-lag estimate. Hermitian symmetry yields $\hat r[-q]=\conj{\hat r}[q]$, and the frequency-domain covariance is
\begin{equation}
    [\hat{\matR}_{\textnormal{f}}]_{i,j}=\hat r[i-j].
    \label{eq:site_specific_toeplitz_covariance}
\end{equation}

\subsubsection{Temporal Covariance Matrix}

For each slot sample, receive antenna, transmission layer, and pilot subcarrier, we interpolate the direct \gls{LS} estimates at the three \gls{DMRS} symbols $t\in\{1,6,11\}$ to all 13 \gls{PUSCH} symbols. We linearly interpolate between adjacent \gls{DMRS} symbols and extrapolate beyond $t=11$ using the estimates at $t\in\{6,11\}$. Stacking channel estimates from all thirteen \gls{OFDM} symbols into vectors and averaging their outer products yields the temporal second moment.

Let $\matA\in\reals^{13\times3}$ contain the corresponding interpolation and extrapolation weights. Assuming independent \gls{LS} estimation errors with variance $\hat{\sigma}_{\textnormal{e}}^2$ at the three \gls{DMRS} positions, the temporal error covariance is
\begin{equation}
    \hat\matSigma_{\textnormal{t}}
    =\hat{\sigma}^2_{\textnormal{e}}\matA\tp{\matA},
\end{equation}
where $\hat{\sigma}_{\textnormal{e}}^2$ is estimated from the processed noise observations. Subtracting $\hat\matSigma_{\textnormal{t}}$ from the temporal second moment yields $\hat{\matR}_{\textnormal{t}}$.

Lastly, we ensure that all estimated covariance matrices are positive semidefinite by setting negative eigenvalues to zero.

\subsection{Receiver Adaptation Results}

Due to the high runtime of Sionna's \gls{LMMSE} channel estimator, the following evaluations use smaller test datasets than the evaluations in \fref{sec:receiver_finetuning_results}. Therefore, the absolute dataset \gls{BLER} values should be compared only with care. Comparisons with the \gls{NRX}, the \gls{MDX}, and the MMSE Reference Rx additionally involve different \gls{LDPC} decoder algorithms.

\subsubsection{Results with Non-Iterative Receivers}

\begin{figure}[tp]
    \centering
    \includegraphics[width=0.9\linewidth]{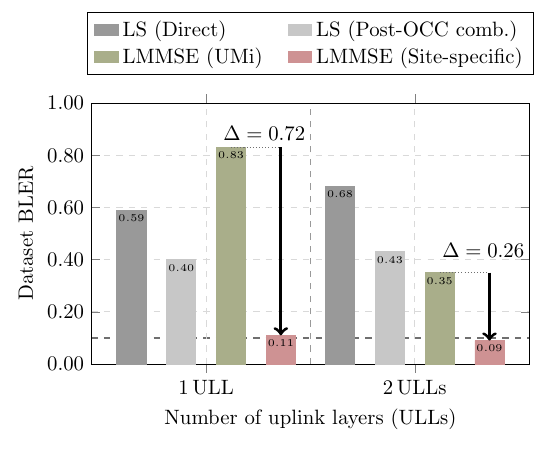}
    \caption{Dataset \gls{BLER} performance of the non-iterative \gls{LMMSE} receiver using linearly interpolated direct or post-OCC-combined \gls{LS} channel estimates or \gls{LMMSE} channel estimates based on synthetic UMi or site-specific covariance estimates. Results use the single-layer (left) and dual-layer (right) small-laboratory datasets from June 2026.}
    \label{fig:lmmse-cov-mixed-mcs-dual-layer}
\end{figure}

\fref{fig:lmmse-cov-mixed-mcs-dual-layer} compares direct and post-OCC-combined \gls{LS} (both use linear interpolation) with \gls{LMMSE} channel estimation in a non-iterative receiver with \gls{LMMSE} \gls{MIMO} detection and $N_{\text{MP}}=12$ \gls{LDPC} \gls{MP} decoding iterations. Post-OCC combining improves the \gls{LS} baseline under both single- and dual-layer transmission due to noise averaging.
The covariance estimates obtained from synthetic UMi channels degrade the single-layer receiver even compared with direct \gls{LS} estimates but improve upon both \gls{LS} configurations under dual-layer transmission.
The single-layer measurements use a lower \gls{PUSCH} \gls{SNR} target than the dual-layer measurements (5\,dB instead of 12\,dB), at which the \gls{LMMSE} estimator relies more strongly on the assumed channel statistics and is therefore more sensitive to covariance mismatch.
Using covariance matrix estimates from site-specific rather than synthetic data reduces the absolute dataset \gls{BLER} by 0.72 and 0.26 for single- and dual-layer transmission, respectively, and closely matches the MMSE Reference Rx.

\subsubsection{Results with Iterative Receivers}

\begin{figure}[tp]
    \centering
    \includegraphics[width=\linewidth]{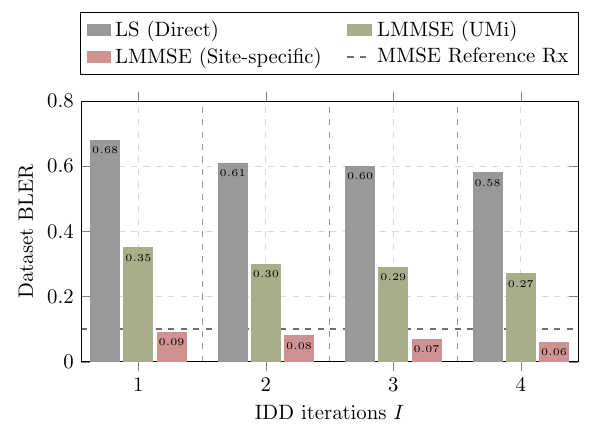}
    \caption{Dataset \gls{BLER} performance versus the number $I$ of MMSE-PIC detection stages using direct \gls{LS} channel estimates or \gls{LMMSE} channel estimates based on synthetic UMi or site-specific covariance estimates. All configurations use $N_{\text{MP}}=12$ \gls{LDPC} \gls{MP} decoding iterations and are evaluated on the small-laboratory dataset from June 2026 with dual-layer transmission.}
    \label{fig:idd-lmmse-cov-lab-dual-layer}
\end{figure}

\fref{fig:idd-lmmse-cov-lab-dual-layer} compares direct \gls{LS} channel estimation with linear interpolation and \gls{LMMSE} channel estimation in an \gls{IDD} receiver for a varying number $I$ of MMSE-PIC \gls{MIMO} detection stages. Each detection stage is followed by $N_i=N_{\text{MP}}/I$ \gls{LDPC} \gls{MP} decoding iterations, which keeps the receiver's total number of decoding iterations fixed at $N_{\text{MP}}=12$~\cite{Wiesmayr2022}. Increasing $I$ reduces the dataset \gls{BLER} with every channel estimator.
\gls{LMMSE} channel estimation with the UMi covariance estimates substantially improves upon direct \gls{LS} channel estimation, while the site-specific covariance estimates provide an additional reduction. The site-specific configuration already slightly outperforms the MMSE Reference Rx without iterative detection at $I=1$ and reduces the dataset \gls{BLER} further from 0.09 to 0.06 at $I=4$. As this is the lowest dataset \gls{BLER} observed throughout this work, this result indicates that channel estimation quality plays a crucial role in overall receiver performance. Furthermore, site-specific parameter adaptation yields even larger dataset \gls{BLER} improvements than site-specific finetuning in \fref{sec:receiver_finetuning_results} and \cite{baytekin2026site}.

\section{Conclusions}

We have shown that site-specific finetuning improves all three evaluated tunable receiver architectures on measured \gls{5GNR} data. The gains are substantial for the two neural receiver architectures---the model-driven \gls{MDX} and fully-tunable \gls{NRX}---but marginal for the model-based \gls{DUIDD} receiver, which has limited tunability.
One \gls{NRX} finetuned on both single- and dual-layer data supports both configurations with only minor performance losses compared with using two separate \glspl{NRX}, one finetuned only on single-layer data and the other only on dual-layer data.
Finetuning remains effective across measurement campaigns separated by more than half a year.
Overall, our results demonstrate that site-specific finetuning improves real-world receiver performance across architectures, \gls{ULL} configurations, and measurement campaigns without increasing runtime complexity.

We have also shown that channel-estimation quality plays a crucial role in overall receiver performance---possibly even more than the choice of the data detection algorithm. Site-specific parameter adaptation through site-specific covariance estimation enables \gls{LMMSE} channel estimation to substantially outperform its counterpart using covariance matrices estimated from synthetic UMi channels.
Combined with this site-specific channel estimator, \gls{IDD} achieves the lowest dataset \gls{BLER} reported in this study and outperforms the MMSE Reference Rx despite using a flooding min-sum decoder rather than pyAerial's layered normalized min-sum decoder.

\bibliographystyle{IEEEtran}

\end{document}